\documentclass[aps,prb,superscriptaddress, twocolumn]{revtex4-1}

\usepackage{graphicx}
\usepackage{epsfig}
\usepackage{units}
\usepackage{color}
\usepackage{textcomp}
\usepackage{amsmath}
\usepackage{amssymb}
\usepackage{hyperref}

\begin{document}

\title{Transport of Spin Qubits with Donor Chains under Realistic Experimental Conditions}

\author{Fahd A. Mohiyaddin}
\affiliation{Centre for Quantum Computation and Communication Technology, School of Electrical Engineering and Telecommunications, UNSW Australia,  Sydney NSW 2052, Australia}

\author{Rachpon Kalra}
\affiliation{Centre for Quantum Computation and Communication Technology, School of Electrical Engineering and Telecommunications, UNSW Australia,  Sydney NSW 2052, Australia}

\author{Arne Laucht}
\affiliation{Centre for Quantum Computation and Communication Technology, School of Electrical Engineering and Telecommunications, UNSW Australia,  Sydney NSW 2052, Australia}

\author{Rajib Rahman}
\affiliation{Network for Computational Nanotechnology, Purdue University, West Lafayette, IN 47907, USA}

\author{Gerhard Klimeck}
\affiliation{Network for Computational Nanotechnology, Purdue University, West Lafayette, IN 47907, USA}

\author{Andrea Morello}
\affiliation{Centre for Quantum Computation and Communication Technology, School of Electrical Engineering and Telecommunications, UNSW Australia,  Sydney NSW 2052, Australia}

\date{\today}

\begin{abstract}
The ability to transport quantum information across some distance can facilitate the design and operation of a quantum processor. One-dimensional spin chains provide a compact platform to realize scalable spin transport for a solid-state quantum computer. Here, we model odd-sized donor chains in silicon under a range of experimental non-idealities, including variability of donor position within the chain. We show that the tolerance against donor placement inaccuracies is greatly improved by operating the spin chain in a mode where the electrons are confined at the Si-SiO$_2$ interface. We then estimate the required timescales and exchange couplings, and the level of noise that can be tolerated to achieve high fidelity transport. We also propose a protocol to calibrate and initialize the chain, thereby providing a complete guideline for realizing a functional donor chain and utilizing it for spin transport.
\end{abstract}

\maketitle

Among the leading physical platforms for the practical implementation of quantum computers, donor spins in silicon \cite{Kane1998n} provide extremely long coherence times \cite{Tyryshkin2012nm, Saeedi2013s} combined with the compatibility with industry-standard fabrication techniques. The last five years have witnessed several experimental milestones in the quest to build a prototype of a donor-based silicon quantum computer. The essential operations of reading out and controlling the spin state of both the electron and nuclear spins of a single implanted $^{31}\mathrm{P}$ donor were demonstrated in a gated nanostructure \cite{Morello2010n, Pla2012n, Pla2013n}. The spin coherence times of the donor electron and nuclear spin qubits in functional nanostructures reached 0.5 s and 30 s, respectively, with state-of-the-art material purification and advanced filtering techniques \cite{Muhonen2014nn}. The scale up of these devices has remained a challenge, although important advancements have been made. Exchange-coupled donor pairs in silicon have been observed \cite{Dehollain2014prl, Gonzalez2014nl} and a two-qubit logic gate has been demonstrated with quantum dots in a similar nanostructure \cite{Veldhorst2015N}. Fabrication based on scanning tunneling microscope (STM) lithography allows for donor incorporation with near-atomic precision \cite{Fuechsle2012nn}.

Beyond the one- and two-qubit logic gates, which can be achieved using short-range interactions and global control fields, the construction of a large-scale quantum computer can greatly benefit from the ability to transport the qubit states across large distances. Even in dense architectures such as the surface code, it is known that long-distance links can help achieving exceptionally high fault-tolerant thresholds \cite{Nickerson2013nc}. Moreover, they can simplify the layout of a quantum processor by allowing extra space between the physical qubits to accommodate control electronics and other components.

Several proposals outline how the spin-carrying electron itself can be transported, whether by shuttling its confinement potential \cite{Taylor2005np}, or by adiabatic passage \cite{Greentree2004prb, Rahman2009prba}. Other schemes involve spin-to-spin coupling between electrons such that transport is essentially achieved via a SWAP operation. This may come from direct exchange coupling, magnetic dipolar interaction \cite{Hill2005prb, Hill2015sa}, electric dipoles \cite{Tosi2015arxiv}, or a coupling mediated via an intermediate quantum dot \cite{Srinivasa2015prl},  a ferromagnet \cite{Trifunovic2013prx} or a resonant cavity \cite{Tosi2014aipadv, Hu2012prb, Tosi2015arxiv}.

\begin{figure}[t!]
\begin{center}
\includegraphics[width=\columnwidth, keepaspectratio = true]{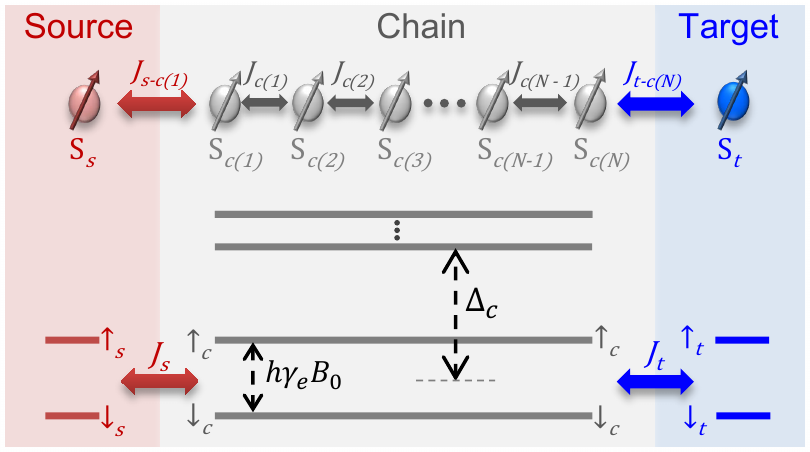}
\caption{(Color online) Schematic of a spin chain (gray) where the spins at the edges (S$_{c(1)}$ and S$_{c(N)}$) are exchange-coupled to source and target qubits (S$_s$ and S$_t$). When $\Delta_c \gg h\gamma_eB_0$, the chain forms an extended qubit, allowing spin transport from the source to the target. The magnitude of the effective coupling between the chain and source qubits, $J_s$, approaches $J_{s-c(1)} / \sqrt{N}$ for large $N$ \cite{Friesen2007prl}.}
\label{fig:chain_intro}
\end{center}
\end{figure}

One-dimensional spin chains have been proposed as a compact medium to couple distant spin qubits \cite{Friesen2007prl, Oh2010prb, Oh2012pra, Oh2013prb}. If the number of spins in the chain is odd and the spin-spin interactions within the chain are very strong, the chain effectively behaves like an extended spin-$1/2$ qubit \cite{Ansatz1931zfp, Meier2003prl}. Therefore, the chain can provide a link between qubits coupled to its ends, as shown in Figure \ref{fig:chain_intro}. A spin state can be transported from a source qubit to the distant target through (i) sequential SWAP operations \cite{Friesen2007prl} or (ii) the adiabatic protocol described by Oh \textit{et al.} \cite{Oh2013pra}.

Earlier theoretical work on spin chains \cite{Friesen2007prl, Oh2010prb, Oh2012pra, Oh2013prb} has given general guidelines on design rules and operation schemes for their use in quantum information processing. However, the practical implementation of a spin chain requires a system-specific appraisal of its physical properties and parameters, and a careful analysis of the manufacturability, error rate and speed of operation under realistic experimental conditions. Here we provide an extensive analysis of the feasibility of spin chains based on donors in silicon.

The paper is organized as follows. In Section \ref{sec:operation_donor_chain}, we outline the requirements for a donor chain to form the desired extended qubit and address the precision with which the donors need to be positioned. In Section \ref{sec:Protocols}, we detail a protocol to calibrate the chain for transport. Finally, in Section \ref{sec:transport}, we assess the fidelity of the adiabatic transport protocol \cite{Oh2013pra} in the presence of magnetic and electrical noise, and with limited tunability of exchange couplings. Our results therefore provide a complete guideline for realizing a functional donor chain and utilizing it for spin transport.

\section{Construction of a Donor Chain}
\label{sec:operation_donor_chain}

We begin our analysis by considering a chain solely made up of electrons, and will only include the donor nuclei later in Section \ref{sec:chain_nucleus}. The spin Hamiltonian for an odd number $N$ of electrons with nearest-neighbor exchange couplings $J_{c(i)}$ in an externally applied static magnetic field $B_0$ is:

\begin{equation}
\label{eq:electron_chain_Hamiltonian}
H_{c_e} = h\gamma_e B_0 \sum\limits_{i=1}^N  \frac{\sigma_{z, c(i)}}{2} + \sum\limits_{i=1}^{N-1}  J_{c(i)}  \frac{\mathrm{\boldsymbol{\sigma}}_{c(i)}}{2} \cdot \frac{\mathrm{\boldsymbol{\sigma}}_{c(i + 1)}}{2},
\end{equation}

where $\gamma_e \approx 28$ GHz/T is the gyromagnetic ratio of the electron. $\boldsymbol{\sigma}_{c(i)}$ is the vector Pauli operator for the $i^\mathrm{th}$ electron spin in the chain and $\sigma_{z,c(i)}$ is the Pauli $z$-operator for the $i$-th spin, with $z$ defined as the direction of the external field $B_0$. For example, for a chain with $N = 3$, $\sigma_{z,c(2)} = I_2 \otimes \sigma_z \otimes I_2$, where $\sigma_z$ is the Pauli  $z$-matrix, and $I_2$ is the $2 \times 2$ identity matrix.

The energy spectrum of an odd-number chain consists of a low-energy doublet of states, separated from the nearest excited states by a gap $\Delta_c$ (see Figure \ref{fig:chain_intro}), that depends on the intra-chain exchange interaction strengths $J_{c(i)}$ . The chain can be treated as an effective two-level system, i.e. a spin-1/2 qubit, under the condition \cite{Friesen2007prl}

\begin{equation}
	\Delta_c  \gg h\gamma_eB_0
	\label{eq:chain_criterion}
\end{equation}

where $h\gamma_eB_0$ is the Zeeman splitting of the ground doublet. Therefore we label the two lowest energy states of the chain as $\lvert\uparrow_{c}\rangle$ and $\lvert \downarrow_{c}\rangle$. Assuming $J_{c(i)} = J_c$ $\forall$ $i$, $\Delta_c$ is given by \cite{Meier2003prl}

\begin{equation}
\Delta_c \approx \frac{J_c \pi^2}{2N}
\label{eq:chain_energy_gap}
\end{equation}

Notice that $\Delta_c$ is inversely proportional to the number of spins in the chain. If the intra-chain exchange couplings are not all equal, $\Delta_c$ needs to be calculated from the numerical diagonalization of Equation \ref{eq:electron_chain_Hamiltonian}. One typically finds that, with inhomogeneous $J_{c(i)}$, $\Delta_c$ is mostly limited by the weaker couplings within the chain.  Regardless of the details of the intra-chain couplings, the chain will function as an extended qubit provided the condition in Equation \ref{eq:chain_criterion} is satisfied. The key point in practice is that the chain serves as an effective spin-1/2 as long as $J_{c(i)} \gg h\gamma_eB_0$  $\forall$ $i$.

Two factors influence the choice of external magnetic field $B_0$. In donor systems, $B_0$ serves the purpose of disentangling the electron and the nuclear spins, which are coupled by the hyperfine interaction $A$. In the example of $^{31}$P in silicon, the hyperfine coupling is $A/h \approx$ 117 MHz \cite{Feher1959pr}. The eigenstates of the $^{31}$P spin Hamiltonian are approximate tensor products of the electron and nuclear states \cite{Steger2011jap} provided $h\gamma_eB_0 \gg A$, thus requiring a minimum $B_0$ $\sim 0.1$ T. Furthermore, the readout of a single donor spin based upon spin-to-charge conversion\cite{Morello2010n} requires that the Zeeman energy $h\gamma_e B_0$ far exceed the thermal energy $k_{\rm B} T$, where $k_{\rm B}$ is the Boltzmann constant and $T$ is the temperature. As $T$ is typically $\sim 100$ mK, a minimum $B_0$ $\sim 1$ T is thus required for high-contrast qubit readout. This value of $B_0$ sets a challenging requirement for the minimum $J_{c(i)}$ in the chain according to Equation \ref{eq:chain_criterion}. For the example of a 7-electron chain, the minimum $J_{c(i)}/h$ must be $> 400$ GHz to satisfy $\Delta_c > 10h\gamma_eB_0$.

\subsection{Accuracy of donor placement and chain operation mode}
\label{sec:donor_placement_accuracy}

The exchange interactions between donor electrons are extremely sensitive to the position of the donor atoms \cite{Koiller2002prl}. Therefore, donor placement accuracy and/or tunability is of paramount importance for quantum devices that exploit the exchange interaction for their functionality. Broadly speaking, there are two methods to controllably incorporate donors within a silicon crystal. The STM method allows near-atomic precision in the placement of the donors \cite{Fuechsle2012nn}, but is not entirely deterministic in the number of donors that end up being incorporated at each location \cite{}, potentially leading to large uncertainty in the actual exchange couplings. Additionally, the low thermal budget in the STM method complicates the growth a high-quality insulating oxide close to the plane that contains the donor, and has so far hindered the ability to electrostatically control STM-incorporated donors through metal gates on the top.

Alternatively, donor atoms can be introduced using the industry-standard ion implantation technique, augmented with methods that allow the counting of each individual ion that enters the substrate \cite{Jamieson2005apl}. Counted single-ion implantation thus overcomes the uncertainty in donor number, but comes at the price of larger inaccuracy in the final location of each implanted donor \cite{Ziegler2010nimb}. However, recent work has shown impressive placement accuracy with a technique where the ions are first cooled and counted inside an ion trap and then accurately focused onto the silicon chip \cite{Jacob2015arXiv}. For a spin chain, variability in $J_{c(i)}$ can diminish $\Delta_c$ past the point where the chain no longer forms a well-defined two-level system. To determine the required positioning accuracy, we must first calculate the exchange interaction between two donors as a function of their separation.

\begin{figure}[t!]
\begin{center}
\includegraphics[width=\columnwidth, keepaspectratio = true]{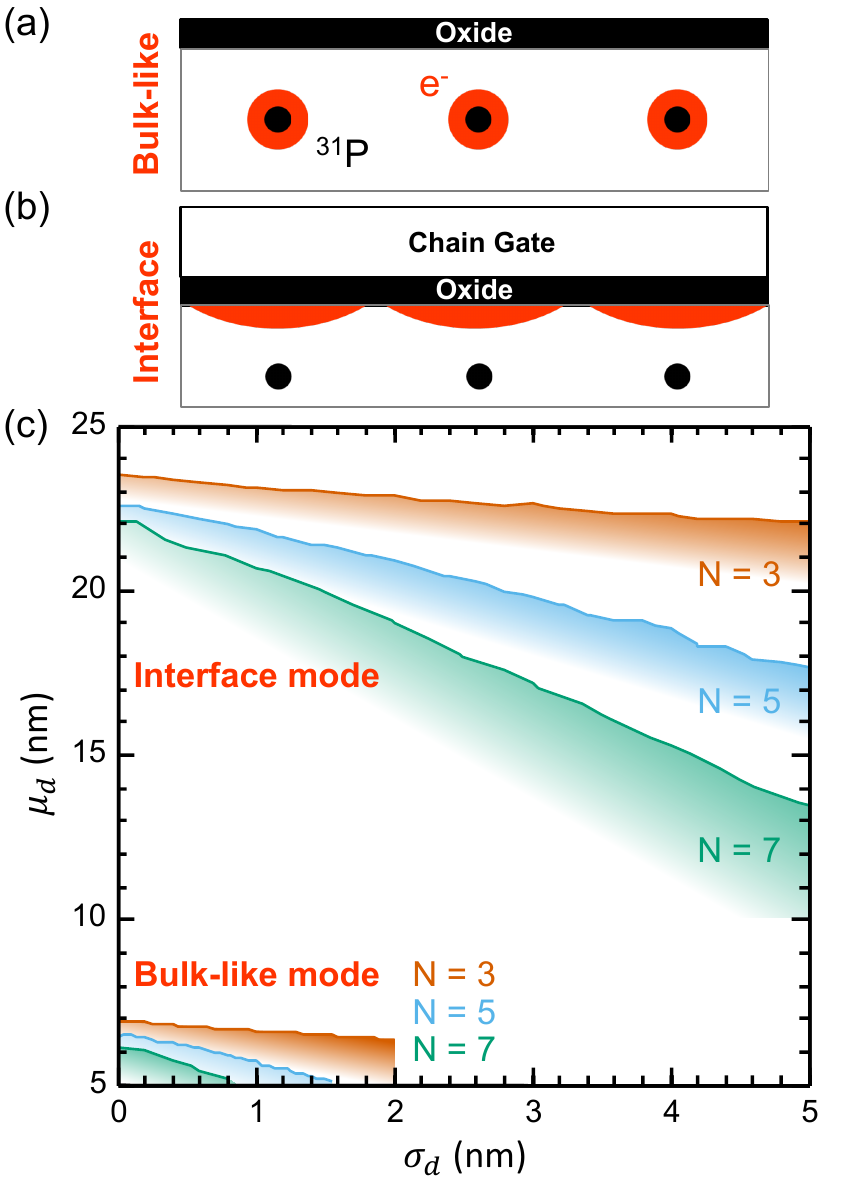}
\caption{(Color online) (a)-(b) Schematics of electron orbital wavefunctions in a three-donor chain operated in the (a) bulk-like and (b) interface modes. The electrons and $\mathrm{^{31}P}$ nuclei are represented in red and black, respectively. (c) Targeted inter-donor separation $\mu_d$ to obtain at least 90\% yield, as a function of donor placement error $\sigma_d$. Chains are simulated for the bulk-like and interface modes for $N =$ 3, 5 and 7, with $B_0 = 1$~T. Pulling electrons to the Si-SiO$_2$ interface with a chain gate increases the robustness of the chain to donor placement inaccuracies.}
\label{fig:chain_yield}
\end{center}
\end{figure}

A key point we wish to raise here is that, provided one has the ability to place an electrostatic gate above the donors, the spin chain can be operated in two distinct modes which have significantly different exchange couplings as a function of inter-donor separation. The first mode is in the absence of an electric field, where the chain electrons are confined to their respective donor nuclei, as they are in bulk silicon. We will refer to this operating regime as the `bulk-like mode,' as illustrated in Figure \ref{fig:chain_yield}(a). The second mode of operation is where a metallic gate located above the chain donors is used to pull the electrons to the Si-SiO$_2$ interface \cite{Calderon2006prl}. For a range of donor depths, 5-20 nm, this `chain gate' can create a vertical electric field sufficient to ionize the donors, while the electrons are still laterally confined by the Coulomb potential of their respective nuclei \cite{Mohiyaddin2014thesis}. We will refer to this operating regime as the `interface mode,' as illustrated in Figure \ref{fig:chain_yield}(b). The calculation of the exchange coupling as a function of inter-donor separation for the two operating modes is given in Appendix \ref{appendix:exchange_calculations}.

On that basis, we proceed to calculate the likelihood that a donor chain fails to satisfy the criterion in Equation \ref{eq:chain_criterion}, typically because enough of the exchange couplings in the chain, $J_{c(i)}$, have become too weak. We perform a Monte Carlo analysis considering chains with $N=$ 3, 5 and 7, with targeted inter-donor separation $\mu_d$. For simplicity, we restrict our analysis to donors placed along the [100] crystallographic plane, where the exchange coupling follows a smooth exponential decay with donor separation \cite{Note1}. An error is introduced in the donor positions along the direction of the chain, following a normal distribution with standard deviation $\sigma_d$. For each chain simulated, we numerically solve the $N$-electron Hamiltonian $H_{c_e}$ to determine the energy separation $\Delta_c$, which determines whether or not the chain functions as an extended qubit. We define the `yield' as the  proportion of chains that satisfy the condition $\Delta_c > 10h\gamma_eB_0$ after performing 10,000 simulations.

Figure \ref{fig:chain_yield} shows the 90\% yield contours for both the bulk-like and interface modes as a function of $\mu_d$ and $\sigma_d$ when $B_0 = 1$ T, for $N=$ 3, 5 and 7. Some clear trends can be identified. The values of $\mu_d$ at $\sigma_d=0$ nm for the six curves in the plot are different. Comparing chains operated in the same mode, we see that the initial $\mu_d$ decreases with increasing $N$, as $\Delta_c$ is inversely proportional to $N$ (Equation \ref{eq:chain_energy_gap}). The allowed separation for the interface mode is also much greater than that of the bulk-like mode. This is due to the greater lateral spread of the orbital wavefunction of the interface electron compared to the donor-bound electron \cite{Calderon2006prl, Mohiyaddin2014thesis}, which leads to a significant enhancement in the exchange couplings $J_{c(i)}$ (refer to Figure \ref{fig:NEMO_exchange} in Appendix \ref{appendix:exchange_calculations}). As $\sigma_d$ increases, $\mu_d$ needs to be reduced to ensure that all the intra-chain exchange couplings are sufficiently large to satisfy the condition in Equation \ref{eq:chain_criterion}. The slope is steeper for chains with a greater number of donors as, for a given $\sigma_d$, there is a greater chance of two adjacent donors having too weak an exchange coupling.

Figure \ref{fig:chain_yield} can be used to determine the chain length achievable given the donor positioning uncertainty for the fabrication process used. For example, with $\sigma_d = 2$~nm, a chain with 3 donors operated in the bulk-like mode is limited to $\mu_d \approx$  6.5 nm, yielding a chain length of only 13 nm. In contrast, much longer chains can be realized when operated in the interface mode. The same uncertainty of 2 nm allows for chains with 3, 5 and 7 donors to have total lengths of $\sim 45$ nm, $\sim 85$ nm and $\sim 115$ nm, respectively. We note that the fabrication overhead to implement an interface-mode chain is minimal, since it only requires one global gate above the entire chain. An equivalent `interface mode' spin chain could be obtained by fabricating a line of electrostatically-defined quantum dots \cite{Lim2009apl,Veldhorst2015N}, but it would come at the cost of fabricating at least one or two individual gates per dot.

In the above calculations, we only considered uncertainty in donor placement along the crystallographic [100] axis. A more general treatment with positioning errors in all directions would be desirable, but is computationally impractical. In the bulk-like mode, misalignments of the donor position away from the [100] axis can severely modify $J_{c(i)}$. This is due to interference between the six ($\mathrm{k_{\pm x}, k_{\pm y}, k_{\pm z}}$) valley components of the donor electron wavefunctions in silicon \cite{Koiller2002prl}. On the other hand, the electron wavefunctions in the interface mode are composed of only the $\mathrm{k_{+z}}$ and $\mathrm{k_{-z}}$ valleys\cite{Saraiva2009prb}, thus removing the valley interference for donors confined to the plane perpendicular to the [001] direction. However, valley interference can still modulate the exchange coupling in the presence of step edges at the Si-SiO$_2$ interface.

Overall, we consider the interface mode to be the preferred mode operation for a spin chain, due to its superior robustness against donor placement inaccuracy, and the ability to build much longer chains with the same donor number as compared to the bulk mode.

\subsection{Source and Target Donor Qubits}
\label{sec:chain_nucleus}

We now include the source and target donor qubits, with their electron spins exchange-coupled to either end of the chain by $J_{s-c(1)}$ and $J_{t-c(N)}$, respectively. Provided $J_{s-c(1)}$ and $J_{t-c(N)} \ll J_c$, the effective exchange coupling between the source (or target) and chain qubits is given by $J_s \approx  J_{s-c(1)}/\sqrt{N}$ (or $J_t \approx  J_{t-c(N)}/\sqrt{N}$) \cite{Friesen2007prl}. Given realistic values of $J_c/h$ of order 100 GHz, the maximum value of $J_s/h$ and $J_t/h$ would be $\sim$ 10 GHz. The source, chain and target electron qubits can then be mapped on to the following Hamiltonian.

\begin{equation}
\begin{split}
\label{eq:Hamiltonian_source_chain_target}
H_{s-c-t} = \epsilon_{s} \frac{\sigma_{z,s}}{2} + \epsilon_{c} \frac{\sigma_{z,c}}{2} +  \epsilon_{t} \frac{\sigma_{z,t}}{2} + \\ J_{s} \frac{\boldsymbol{\sigma}_s}{2}\cdot\frac{\boldsymbol{\sigma}_c}{2} + J_{t}\frac{\boldsymbol{\sigma}_c}{2}\cdot\frac{\boldsymbol{\sigma}_t}{2},
\end{split}
\end{equation}

where $\boldsymbol{\sigma}_{i}$ is the Pauli operator with $z$ component $\sigma_{z,i}$. $\epsilon_i$ is the energy splitting between the qubit states, where the subscript $i$ denotes the source, chain or target.

We assume that the source and target donors are operated in the bulk-like mode, where the electron and nuclear spins are coupled by the hyperfine interaction $A$, and that $h\gamma_e B_0 \gg A$ to ensure that the electron-nuclear eigenstates of source and target qubits are disentangled. In this regime, the hyperfine interaction simply modifies the electron qubit splitting by an amount dependent on the nuclear spin state. The latter is known to remain unchanged for several minutes \cite{Pla2013n} unless forcibly modified by the application of radio-frequency excitations. The source and target qubit splitting, $\epsilon_s$ and $\epsilon_t$, are equal to $h\gamma_e B_0 + A/2$ or $h\gamma_e B_0-A/2$ when the nucleus is in the $\lvert\Uparrow\rangle$ state or $\lvert\Downarrow\rangle$ state, respectively. In the analysis below, we set $\epsilon_{s} = \epsilon_{t}$, which can be realized by preparing the nuclear spins of the source and target donors in the same state, and tuning their hyperfine couplings with local electrostatic gates \cite{laucht2015sciadv} until they acquire identical values.

We address the effect of the nuclear spins on a chain operated in the bulk-like mode in Appendix \ref{appendix:bulk_like_mode}. For a chain in the interface mode, however, the hyperfine coupling between the electron and nuclear spins is zero, since the electron wavefunctions do not overlap with those of the nuclei. Therefore, the energy separation $\epsilon_c$ between the $\lvert\uparrow_c\rangle$ and $\lvert\downarrow_c\rangle$ chain qubit states is simply equal to $h\gamma_e B_0$.

We define the energy detuning between the source and chain qubits as $\Delta B_z = \lvert \epsilon_{s} - \epsilon_{c} \rvert$. In the case of an interface-mode chain, $\Delta B_z = A/2$ regardless of the source qubit nuclear state. We use the notation $\Delta B_z$ to highlight that this detuning coincides with the energy difference between the $\lvert\uparrow_s\downarrow_c\rangle$ and $\lvert\downarrow_s\uparrow_c\rangle$ source-chain states, and has the same physical origin as the energy difference between the $\lvert\uparrow\downarrow\rangle, \lvert\downarrow\uparrow\rangle$ states in a singlet-triplet qubit. Here, however, $\Delta B_z$ does not depend on the polarization of a large bath of nuclear spins, as would be the case in a double quantum dot system, but simply arises from the fact that one qubit (the source or target) is coupled to a single nuclear spin, while the other (the chain) is not. Indeed, recent experiments have shown the potential of this type of donor-dot hybrid systems to realize singlet-triplet qubits with robust values of $\Delta B_z$ \cite{HarveyCollard2015arXiv, Urdampilleta2015prx}.

\section{Calibration of the System}
\label{sec:Protocols}

Spin transport across the chain requires control over the exchange couplings $J_s$ and $J_t$, as will be described in Section \ref{sec:transport}. This control may come from tuning the tunnel barriers between the donor electrons directly \cite{Wellard2003prb}, or detuning their respective electrochemical potentials \cite{Wang2015arxiv}, with gate electrodes. It is however extremely unlikely, even with atomically precise donor placement, that the magnitude of exchange couplings will match the values targeted during fabrication. Therefore, it will be necessary to first calibrate $J_s$ and $J_t$ to the voltages of the respective gate-electrodes designed to tune them (exchange-gates). For spin transport, the key quantities to record are the minimum and maximum values of the exchange couplings that can be achieved.  In addition, it is also important to measure $\epsilon_i$ for each qubit, as it can vary due to magnetic field inhomogeneities and DC Stark shifts of the electron $g$-factor \cite{Rahman2009prbb} and donor hyperfine interaction \cite{Mohiyaddin2013nl, laucht2015sciadv}.

\begin{figure}[t!]
\begin{center}
\includegraphics[width=\columnwidth, keepaspectratio = true]{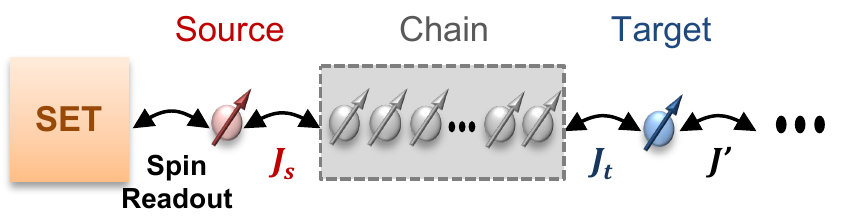}
\caption{(Color online) Schematic of the system used for transport. A source donor is exchange-coupled to a donor chain, which is in turn coupled to a target donor. The target donor is then linked to the remainder of the quantum processor. A single-electron transistor (SET) tunnel-coupled to the source donor serves to initialize and measure the state of the source electron spin qubit.}
\label{fig:Architecture}
\end{center}
\end{figure}

Before providing a calibration protocol, we first introduce the way in which a spin chain might be incorporated into a quantum processor architecture, such as the one presented in Reference \onlinecite{Hollenberg2006prb}. A source donor at the edge of the quantum processor is tunnel-coupled to a single-electron transistor (SET) for initialization and readout of its electron and nuclear spins \cite{Morello2009prb, Morello2010n, Pla2013n}, as illustrated in Figure \ref{fig:Architecture}. This donor is then exchange coupled to a donor chain, which is in turn coupled to a target donor. The target donor is linked to the remaining entities of the processor. Due to layout constraints, it may not be possible to fabricate an SET for every donor qubit in the processor for initialization and readout. We have thus developed a protocol to calibrate and initialize the qubits, regardless of their distance from the edge of the processor.

Our calibration protocol relies on the assumption that any qubit or a pair of qubits can be effectively isolated from the remainder of the processor. For example, when the source donor is being calibrated, one must ensure that the chain does not alter its dynamics.  Similarly, when $J_s$ is being calibrated, the target should not alter the dynamics of the source-chain system. This isolation may be achieved in two steps. We present these steps using the example of isolating the source donor. First, $J_s$ should be minimized by pulsing its exchange-gate appropriately. However, it would be unrealistic to assume that $J_s = 0$. The source qubit, therefore, is not completely separated from the dynamics of the chain.  The second step utilizes $J_t$ to minimize the effect of a non-zero $J_s$.  Maximizing $J_t$, i.e. strongly coupling the chain and target qubits, has the effect of isolating the dynamics of the source qubit. This is because the eigenstates of the system will then approximately be the tensor products of the uncoupled source qubit states with the singlet-triplet states of the chain and target, similar to the case where $J_s=0$. The required ratio of $J_t$ to $J_s$ for this isolation will be quantified in Section \ref{sec:AP_Jmin}. We assume that the above steps are sufficient to isolate any qubit (or pair of qubits) in the processor that is being calibrated.

We begin by determining $\epsilon_s$ of the isolated source donor. The spin state of the source electron is measured and initialized using spin-dependent tunneling to the SET \cite{Morello2010n}. $\epsilon_s$ can then be extracted using the electron spin resonance (ESR) technique outlined in Reference \onlinecite{Pla2012n}.

The next step is to calibrate $J_s$ with its exchange-gate ($J_s$-gate) voltage. For this, the chain is first isolated from the target, by minimizing $J_t$ and maximizing $J'$ (Figure \ref{fig:Architecture}). Then, the ESR spectrum of the source qubit is measured while varying the $J_s$-gate voltages. The exchange coupling $J_s$ modifies the source resonance frequencies and provides a unique `fingerprint' that can be compared to the theoretically calculated ESR spectrum described below, resulting in an accurate map of $J_s$ as a function of $J_s$-gate voltage.

\begin{figure}[t!]
\begin{center}
\includegraphics[width=\columnwidth, keepaspectratio = true]{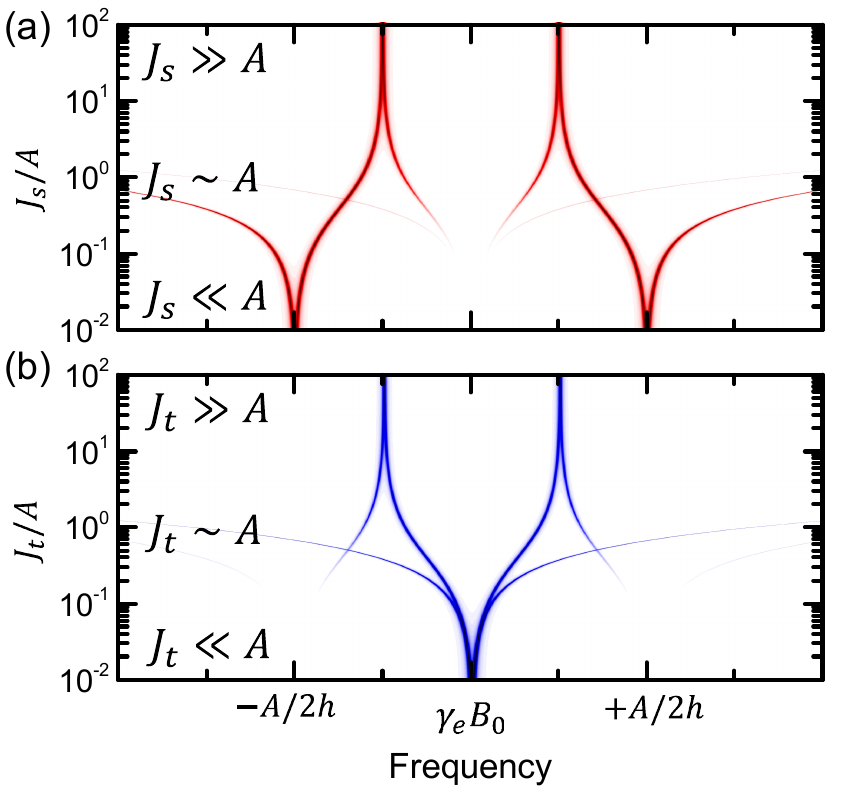}
\caption{(Color online) ESR spectra of a donor electron coupled to a chain qubit, operated in the interface mode. The spectra are shown for the cases where measurement of only (a) the donor electron or (b) the chain qubit is possible. (a) and (b) are used to calibrate $J_s$ and $J_t$, respectively.}
\label{fig:ESRspectrum_source_chain_interface}
\end{center}
\end{figure}

Figure ~\ref{fig:ESRspectrum_source_chain_interface}(a) shows the ESR spectrum of the source electron as a function of $J_s$, calculated by solving for the eigenstates of the Hamiltonian:

\begin{equation}
\begin{split}
H_{s-c(\mathrm{interface})} = h\gamma_eB_0 \left( \frac{\sigma_{z,s}}{2}  + \frac{\sigma_{z,c}}{2} \right) - h\gamma_nB_0\frac{^{\mathrm{nuc}}\sigma_{z,s}}{2} + \\ A \frac{\boldsymbol{\sigma}_s}{2}\cdot\frac{^{\mathrm{nuc}}\boldsymbol{\sigma}_s}{2} + J_s \frac{\boldsymbol{\sigma}_s}{2}\cdot\frac{\boldsymbol{\sigma}_c}{2},
\end{split}
\label{eq:Hamiltonian_source_chain_interface}
\end{equation}

where $^{\mathrm{nuc}}\boldsymbol{\sigma}_s$ is the Pauli operator for the source nucleus with $z$-component $^{\mathrm{nuc}}\sigma_{z,s}$, and $\gamma_n \approx 17.2$ MHz/T is the nuclear gyromagnetic ratio for the $\mathrm{^{31} P}$ donor. To obtain a model ESR spectrum that would match the experiment described, for each value of $J_s$, we weigh all possible transitions between the eigenstates of $H_{s-c(\mathrm{interface})}$ with the product of the transition probability and spin readout contrast of the source qubit ~\cite{Kalra2014prx}. This corresponds to the readout signal available in the experiment, which is the spin of the source electron. The resonance frequencies obtained in the experiment at a particular voltage on the $J_s$-gate corresponds to a horizontal slice in the plot.

We now briefly describe the physics in Figure \ref{fig:ESRspectrum_source_chain_interface}(a). For $J_s \ll A/2$, the ESR spectrum is that of the isolated source donor, where the two hyperfine-split peaks \cite{Pla2012n} correspond to the two possible frequencies of $\epsilon_s/h = \gamma_eB_0 \pm A/2h$. As $J_s$ is increased with its exchange-gate, each of the peaks split into two branches corresponding to the two possible states of the chain qubit, $\lvert\uparrow_c\rangle$ and $\lvert\downarrow_c\rangle$. This splitting is equal to $J_s/h$, allowing its magnitude to be directly obtained in the low $J_s$ ($ < A/2$) regime. As $J_s$ is increased further ($> \Delta B_z =  A/2$), the eigenstates of the coupled source and chain qubits evolve towards the spin-singlet $\lvert S\rangle$ and the triplet states, $\lvert T_0\rangle$, $\lvert T_+\rangle$ and $\lvert T_-\rangle$. The branches that involve the $\lvert{S}\rangle$-like state fade away as their transition probabilities tend to zero. In contrast, the branches that involve the $\lvert{T_0}\rangle$-like state tend towards a frequency that is the average of the isolated source and chain qubit frequencies, i.e. $(\epsilon_s + \epsilon_c)/2h$. This is equal to $\gamma_eB_0 \pm A/4h$ as shown at the top of Figure \ref{fig:ESRspectrum_source_chain_interface}(a), depending on the spin of the source nucleus.

While the low ($ < A/2$) values of $J_s$ can be extracted directly from the ESR spectrum, a different technique is required to estimate them when $J_s > A/2$. In this regime, $J_s$ can be measured using a SWAP-style experiment, as detailed in the sequence below.

\renewcommand{\theenumi}{\roman{enumi}}%
\begin{enumerate}
	\item Initialize the source-chain system in anti-parallel states while $J_s$ is minimized. For this, the chain qubit needs to be read out using a conditional-rotation (CROT) on the source qubit \cite{Kalra2014prx}: with the source initialized in the $\lvert\downarrow\rangle$ state and $J_s$ pulsed to $\ll A/2$, an ESR $\pi$-pulse is applied at the frequency where the source qubit flips only if the chain is in the $\lvert\downarrow\rangle$ state. If the source has not flipped, then the chain has been determined to be in the $\lvert\uparrow\rangle$ state. The source qubit should then be initialized in the opposite state to the chain qubit.
	\item Measure the frequency of exchange oscillations. With the system initialized in anti-parallel states, $J_s$ is pulsed high ($ > A/2$) for a time $\tau$, and then minimized thereafter. The source qubit is then readout to see if it has flipped. This is repeated several times to obtain a flip-probability. The flip-probability can be plotted as a function of $\tau$, and will display exchange oscillations at frequency $J_s$.
\end{enumerate}

The chain qubit splitting $\epsilon_c$ is calibrated by measuring its ESR spectrum, using the fact that its spin state can be read out via CROT of the source qubit in the $J_s \ll A/2$ regime. Similarly, access to the spin state of the chain qubit allows the calibration of $J_t$ to its associated exchange-gate with the same method used to calibrate $J_s$. Figure ~\ref{fig:ESRspectrum_source_chain_interface}(b) shows the ESR spectrum of the chain qubit as a function of $J_t$. This spectrum is obtained by solving the Hamiltonian in Equation \ref{eq:Hamiltonian_source_chain_interface}, but where the source is replaced by the target and the readout contrast is based on the chain qubit instead. In the low $J_t$ regime, the ESR spectrum is that of the isolated chain qubit, whereas the spectra of Figure ~\ref{fig:ESRspectrum_source_chain_interface}(a) and (b) converge in the high $J_s$ or $J_t$ regime.

The above techniques can thus be used recursively to (i) initialize and readout any qubit, (ii) measure $\epsilon_i$ of each qubit, and (iii) calibrate the exchange interaction to its associated gate for any pair of qubits in the processor. As we shall explain below, a particularly important parameter for the operation of the chain is the maximum exchange coupling achievable between chain and qubits, $J_{\rm max}$.

Note that our analysis in this section focused on a donor chain operated in the interface mode. For completeness, we also present the calibration protocol and ESR spectra for the case where the chain is operated in the bulk-like mode in Appendix \ref{appendix:bulk_like_mode_calibration}.

\section{Spin Transport}
\label{sec:transport}

\begin{figure}[t!]
\begin{center}
\includegraphics[width=\columnwidth, keepaspectratio = true]{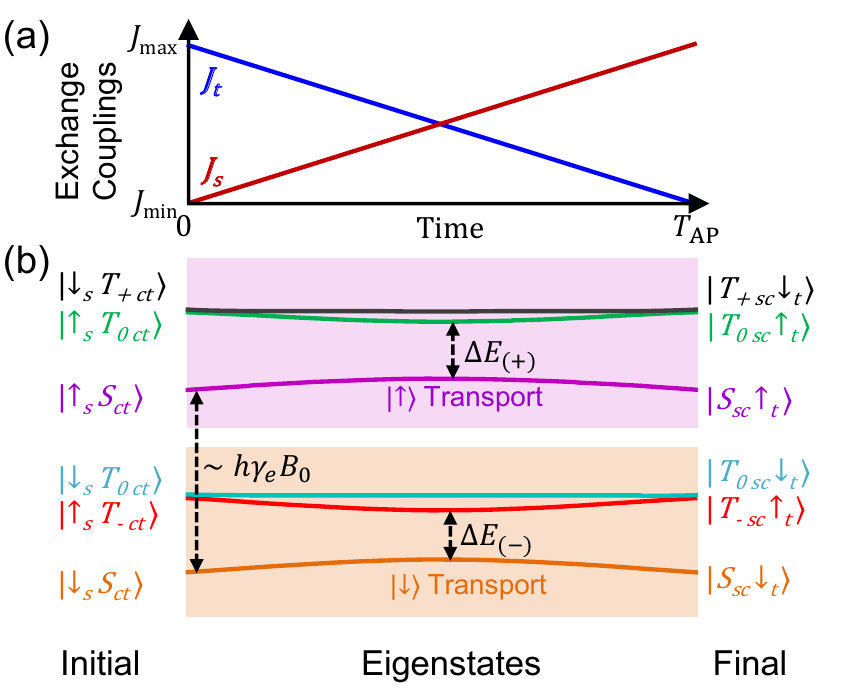}
\caption{(Color online) (a) Pulsing scheme for the adiabatic spin transport protocol. (b) Sketch of the time evolution of the eigenenergies ($y$-axis not to scale). The approximate eigenstates are labeled at the start and end of the protocol. The $\lvert\uparrow\rangle$ and $\lvert\downarrow\rangle$ states are transported via the two labeled adiabatic passages. They belong to two independent blocks of the Hamiltonian as grouped by the shaded boxes.}
\label{fig:APEigStates}
\end{center}
\end{figure}

In this section, we analyze the transport of a source qubit to the target qubit via the chain. For our analysis, we define the transport fidelity as:

\begin{equation}
\label{eq:transport_fidelity}
F = \lvert\langle\Psi_f\vert \Psi_r\rangle\rvert^2,
\end{equation}

where $\lvert\Psi_r\rangle$ is the required final state of the source-chain-target system and $\lvert\Psi_f\rangle$ is the actual final state of the system after transport.

An intuitive method for transporting the source qubit to the target is via sequential SWAP operations \cite{Friesen2007prl}, where the spin state is first transferred to the chain and then to the target. A SWAP operation is achieved by pulsing the exchange coupling $J$ (i.e. $J_s$ or $J_t$) to a value much larger than $\Delta B_z$ (Reference \onlinecite{Kalra2014prx}), for a time $T_{\mathrm{SWAP}}$, such that $\int_0^{T_\mathrm{SWAP}} \left[J(t)/h\right] dt = 0.5$. For example, $T_\mathrm{SWAP}$ needs to be 50 ps for $J(t)/h=10$ GHz. Defining $\bar{J}$ as the mean exchange coupling during a SWAP operation, the transport fidelity is given by

\begin{equation}
\label{eq:Fswap}
F_\mathrm{SWAP} \approx \sin^4{(\pi T_{\mathrm{SWAP}}\bar{J}/h )}.
\end{equation}

While the SWAP operation is fast, $F_\mathrm{SWAP}$ is sensitive to noise in $J$ and timing imperfections. Equation \ref{eq:Fswap} shows that an accuracy of $\bar{J} T_{\mathrm{SWAP}}$ to within 2\% is required to obtain $F_\mathrm{SWAP} > 99$\%. In the example where $J(t)/h =  10$ GHz and $T_\mathrm{SWAP} = $ 50 ps, this translates to a requirement of pulses with picosecond precision. To circumvent this timing constraint, an adiabatic transport protocol robust to pulsing errors was proposed in Reference \onlinecite{Oh2013pra}.

An adiabatic process is one in which the instantaneous eigenstates of the system are modified at a rate much slower than the energy separations between them. The Hamiltonian $H_{s-c-t}$ in Equation \ref{eq:Hamiltonian_source_chain_target} is block diagonal, as explained in Reference \onlinecite{Oh2013pra}. We analyze the adiabatic transport process by starting with $J_s = 0$ and $J_t = J_\mathrm{max}$ at $t=0$. At this point, the eigenstates are the uncoupled source spin state and the chain-target singlet and triplet states, as labeled on the left side of Figure \ref{fig:APEigStates}(b). The source holds the qubit state to be transported, $\lvert \psi_s \rangle$, and the chain and target qubits must be initialized in the singlet state $\lvert S_{ct} \rangle$ (see Section \ref{sec:SingletInitialization} below). The system is thus in a superposition of the $\lvert\uparrow_s S_{ct}\rangle$ and $\lvert \downarrow_sS_{ct}\rangle$ eigenstates, which belong to two independent three-state blocks in the Hamiltonian. These blocks are grouped by the shaded boxes in Figure \ref{fig:APEigStates}(b). Note here that we have omitted the $\lvert\uparrow_s T_{+ct}\rangle$ and $\lvert \downarrow_s T_{- ct}\rangle$ states in the figure as they are in separate blocks of $H_{s-c-t}$ and do not play a role in transport.

Once initialized, the transport protocol is completed by ramping $J_s$ towards $J_{\rm max}$ and $J_t$ towards 0 over a time $T_\mathrm{AP}$, as shown in Figure \ref{fig:APEigStates}(a). The evolution of the eigenenergies as a function of time is shown in Figure \ref{fig:APEigStates}(b), revealing that the $\lvert\uparrow_s\rangle$ and $\lvert\downarrow_s\rangle$ components follow two independent adiabatic passages. The two passages are identical if $\Delta B_z = 0$. At the end of the protocol, the eigenstates are essentially reflections of the $t=0$ states. The source qubit is transported to the target, and the prepared singlet is reflected on to the source and chain qubits.

Prior to estimating the fidelity of the adiabatic protocol (Section \ref{sec:adiabatic_transport}), we will outline a method to initialize the target and chain for transport.

\subsection{Singlet Initialization for Adiabatic Transport}
\label{sec:SingletInitialization}

\begin{figure}[t!]
\begin{center}
\includegraphics[width=\columnwidth, keepaspectratio = true]{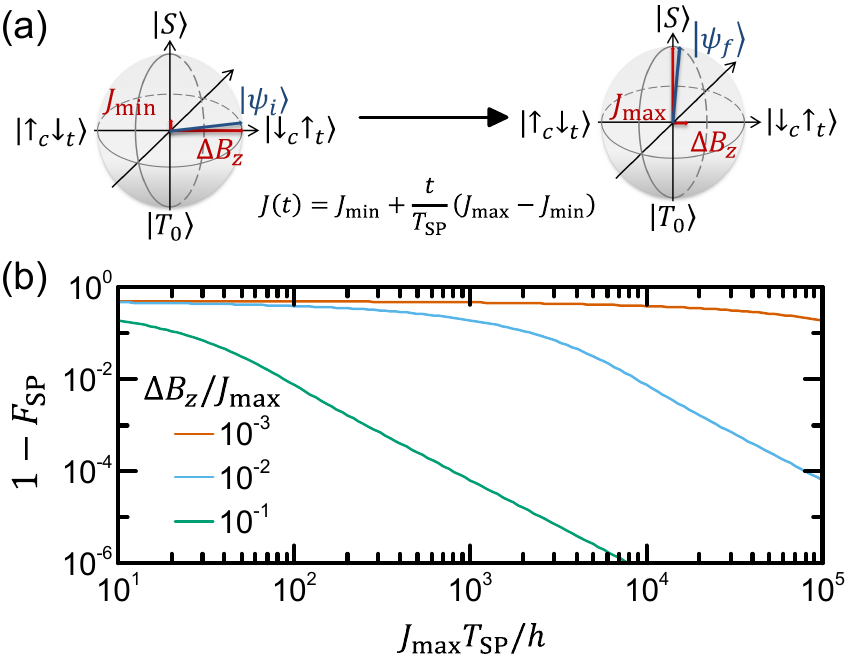}
\caption{ (Color online) (a) Protocol for initializing the chain and target qubits into the near-singlet $\lvert\varphi_{S_{ct}}\rangle$ state of chain and target qubits. (b) Initialization error as a function of the product of $T_\mathrm{SP}$ and $J_\mathrm{max}$, for three values of $\Delta B_z/J_\mathrm{max}$. The fidelities are approximately given by Equation \ref{eq:Fidelity_SP}.}
\label{fig:APSingletInitialization}
\end{center}
\end{figure}

Recall from Section \ref{sec:operation_donor_chain} that the required magnetic field is $\sim 1$ T and the maximum value of $J_s/h$ and $J_t/h$ is $\sim 10$ GHz. Therefore, the ground-state of the chain-target system when $J_t$ is at its maximum is the $\lvert T_{- ct} \rangle$ state, rather than the singlet $\lvert S_{ct}\rangle$ state. This rules out several well-established techniques to initialize two qubits in the singlet state that require it to be their ground state \cite{Petta2005s}. Below we show that we can nevertheless initialize a singlet state by making use of the presence of a finite $\Delta B_z$. The idea is similar to the way in which the $\lvert\uparrow\downarrow\rangle$ or $\lvert\downarrow\uparrow\rangle$ states are initialized in singlet-triplet qubits \cite{Petta2005s}.

We begin with the chain and target qubits in the ground $\lvert T_{-ct}\rangle$ state with minimal $J_s$ and $J_t$, using the techniques described in Section \ref{sec:Protocols}. For ease of explanation, we assume that the nuclei of the source and target donors are both initialized in the $\lvert\Downarrow\rangle$ state. ESR is then used to excite the two qubits to $\lvert\psi_i\rangle$ as labeled in Figure \ref{fig:APSingletInitialization}(a). $\lvert\psi_i\rangle$ is the lower-energy state of the two anti-parallel eigenstates in the low-$J$ regime, and is equal to $\lvert\widetilde{\downarrow_c\uparrow_t}\rangle = \cos(\theta_2)\lvert \downarrow_c\uparrow_t\rangle + \sin(\theta_2)\lvert \uparrow_c\downarrow_t \rangle$, where $\tan(2\theta_2) = J_\mathrm{min}/\Delta B_z$. $J_t$ is then increased adiabatically to $J_\mathrm{max} (\gg \Delta B_z)$ over a timescale $T_{\mathrm{SP}}$ such that the initialized state $\lvert\psi_i\rangle$ evolves to $\lvert\psi_f\rangle$, as shown in Figure \ref{fig:APSingletInitialization}(a).

Observe that the prepared state $\lvert\psi_f\rangle$ is not exactly equal to the singlet state. $\Delta B_z$ modifies the eigenstate of the chain-target system from the exact singlet to a `near singlet' state, $\lvert\varphi_{S_{ct}}\rangle = \cos(\theta_1)\lvert S_{ct}\rangle - \sin(\theta_1)\lvert T_{0 ct} \rangle$, where $\tan(2\theta_1) = \Delta B_z / J_\mathrm{max}$. The effect of the discrepancy between the $\lvert\varphi_{S_{ct}}\rangle$ state and the ideal initial state $\lvert S_{ct}\rangle$ on the adiabatic transport protocol will be addressed later in Section \ref{sec:AP_DBz}. Here, we only focus on the fidelity with which this initialization protocol prepares the $\lvert\varphi_{S_{ct}}\rangle$ state.

The fidelity $F_{\mathrm{SP}}$ is calculated as a function of $T_{\mathrm{SP}}$, the time over which $J_t$ is ramped. $F_{\mathrm{SP}}$ is defined as the squared-projection of the final chain-target state onto the required $\lvert\varphi_{S_{ct}}\rangle$ state. Figure \ref{fig:APSingletInitialization}(b) plots the error $ 1 - F_{\mathrm{SP}}$ as a function of the product $J_\mathrm{max}T_\mathrm{SP}/h$ for $\Delta B_z/J_\mathrm{max} = 10^{-1}$, $10^{-2}$ and $10^{-3}$.  For simplicity we assume $J_\mathrm{min} = 0$. Recall that the adiabaticity of an adiabatic protocol is enhanced when the minimum energy separation between the eigenstates  and/or the duration of the protocol is increased. Therefore the fidelity improves as $T_\mathrm{SP}$ is increased, as long as dephasing can be neglected. This is illustrated in Figure \ref{fig:APSingletInitialization}(b) by considering a fixed $J_\mathrm{max}$ and $\Delta B_z$. In addition, the fidelity improves by increasing $\Delta B_z$, since it determines the minimum energy separation between the two eigenstates in this Bloch sphere during the adiabatic protocol.  This is also observed in the figure by considering a fixed $J_\mathrm{max}$ and $T_\mathrm{SP}$. The expression below, derived in Appendix \ref{appendix:fidelity_calculations}, is an approximation for $F_\mathrm{SP}$ in the adiabatic regime:

\begin{equation}
1 - F_\mathrm{SP} \approx \frac{1}{4\left( J_\mathrm{max}T_\mathrm{SP}/\hbar\right)^2} \left(\frac{J_\mathrm{max}}{\Delta B_z}\right)^4.
\label{eq:Fidelity_SP}
\end{equation}

Figure \ref{fig:APSingletInitialization}(b) shows that high initialization fidelities are achievable with this protocol. For example, an initialization error of $<10^{-4}$ can be obtained for $\Delta B_z/h = 100$ MHz, $J_\mathrm{max}/h = 10$ GHz, and $T_\mathrm{SP} > 10$ $\mu$s. However, we note that the required $T_\mathrm{SP}$ are several orders of magnitude larger than the timescales for the transport protocol, as will be discussed in the next section. Nonetheless, the initialization protocol need only be performed once, since the prepared state can be reused. This is due to the fact that the transport protocol reflects the $\lvert\varphi_{S_{ct}}\rangle$ state on to the source and chain qubits, as illustrated in Figure \ref{fig:APEigStates}(b).

\subsection{Adiabatic transport under realistic experimental conditions}
\label{sec:adiabatic_transport}

We will now investigate the influence of the following experimental parameters on the adiabatic transport protocol: (i) Errors in $T_\mathrm{AP}$ and/or $J_\mathrm{max}$; (ii) $\Delta B_z$ between the chain and source/target qubits; (iii) Limited tunability of the exchange couplings;  (iv) Noise in the qubit energy splittings $\epsilon_i$; (v) Noise in the exchange couplings $J_s$ and $J_t$. We define the transport fidelity $F_\mathrm{AP}$ for transporting a source state $\lvert\psi_s\rangle$ according to Equation \ref{eq:transport_fidelity}. Here, the required final state of the system is $\lvert \Psi_r \rangle = \lvert \varphi_{S_{ct}}\rangle \otimes \lvert \psi_s\rangle$.

\subsubsection{Errors in $J_\mathrm{max}$ and $T_\mathrm{AP}$}
\label{sec:AP_JmaxTAP}

$J_\mathrm{max}$ and $T_\mathrm{AP}$ are the fundamental transport parameters as they determine the degree of adiabaticity of the protocol \cite{Oh2013pra}. To illustrate this, it is instructive to first consider the case where $\Delta B_z = 0$. In this case, the minimum energy separations $\Delta E_{(\pm)}$ between each adiabatic passage and the nearest eigenstate in their blocks are equal to $J_\mathrm{max}/2$ [see  Figure \ref{fig:APEigStates}(b)]. To maintain adiabaticity, we require the transport time $T_\mathrm{AP} \gg h/\Delta E_{(\pm)}$,  and hence the transport fidelity $F_\mathrm{AP}$ is dependent on the product $J_\mathrm{max} T_\mathrm{AP}/h$.

\begin{figure}[t!]
	\begin{center}
		\includegraphics[width=\columnwidth, keepaspectratio = true]{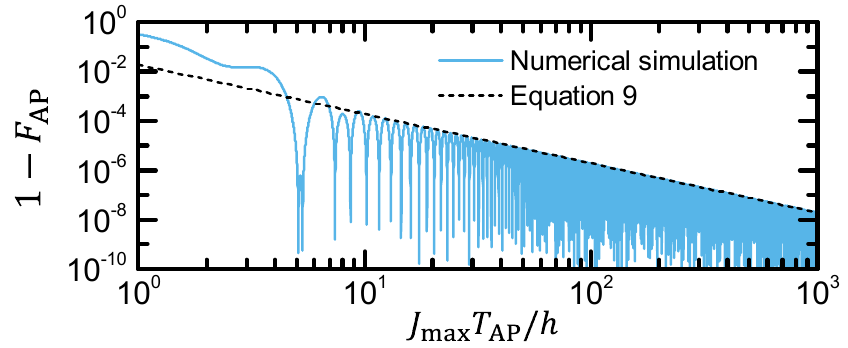}
		\caption{(Color online) Error of the transport protocol as a function of $J_\mathrm{max}T_\mathrm{AP}/h$, assuming $\Delta B_z = 0$ and $J_\mathrm{min} = 0$. The envelope of the transport error in the adiabatic regime is given by Equation \ref{eq:Fidelity_AP_JmaxTAP} (dashed line). The protocol is highly robust to errors in $J_\mathrm{max}$ and $T_\mathrm{AP}$.}
		\label{fig:APFidelity_JmaxTAP}
	\end{center}
\end{figure}

Figure \ref{fig:APFidelity_JmaxTAP} shows the transport error, $1-F_\mathrm{AP}$, as a function of $J_\mathrm{max}T_\mathrm{AP}/h$, obtained from numerical simulations of transporting the $\lvert\psi_s\rangle = (\lvert\uparrow\rangle+\lvert\downarrow\rangle)/\sqrt{2}$ state. We note that fidelities calculated here are independent of the choice of $\lvert\psi_s\rangle$. The resonances indicate points where perfect adiabatic transport is achieved \cite{Oh2013pra}. However, we focus on the envelope to provide a conservative estimate of fidelities. For large $J_\mathrm{max}T_\mathrm{AP}/h$, Reference \onlinecite{Oh2013pra} showed that the envelope of the error is proportional to $1/(J_\mathrm{max}T_\mathrm{AP})^2$.  This envelope can also be obtained analytically (Appendix \ref{appendix:fidelity_calculations}), and is given by

\begin{equation}
1 - F_\mathrm{AP} \approx \frac{3}{3 + \left(2 J_\mathrm{max}T_\mathrm{AP}/\hbar\right)^2}.
\label{eq:Fidelity_AP_JmaxTAP}
\end{equation}

The above equation is also plotted in Figure \ref{fig:APFidelity_JmaxTAP} as the dashed black line. This highlights that the adiabatic protocol is robust to errors in $J_\mathrm{max}$ and $T_\mathrm{AP}$, and low transport errors of $\sim$ $10^{-4}$, $10^{-6}$ and $10^{-8}$ can be achieved for $J_\mathrm{max} T_\mathrm{AP}/h \approx $ 10$^1$, 10$^2$ and 10$^3$, respectively \cite{Oh2013pra}. Therefore, depending on the targeted fidelity, transport times $T_\mathrm{AP} \sim 10 - 100$~ns are required when $J_\mathrm{max}/h \sim 1 - 10$~GHz.

Note that the timescales required for $T_\mathrm{AP}$ are significantly shorter than those required for the singlet initialization protocol time $T_\mathrm{SP}$. This is because the transport time is limited by $J_\mathrm{max}$, while $T_\mathrm{SP}$ is limited by $\Delta B_z \ll J_\mathrm{max}$.

\subsubsection{The effect of $\Delta B_z$}
\label{sec:AP_DBz}

The ratio between $\Delta B_z$ and $J_\mathrm{max}$ also has an important effect on the adiabaticity of the protocol. The energy separations $\Delta E_{(\pm)}$ as a function of $\Delta B_z$ are given by:

\begin{equation}
\frac{\Delta E_\mathrm{(\pm)}}{J_\mathrm{max}} = \frac{1 + \sqrt{9 \pm 8 \frac{\Delta B_z}{J_\mathrm{max}} + 16(\frac{\Delta B_z}{J_\mathrm{max}})^2 }}{8} \pm \frac{\Delta B_z}{2 J_\mathrm{max}} .
\label{eq:APEnergySeparation}
\end{equation}

\begin{figure}[t!]
\begin{center}
\includegraphics[width=\columnwidth, keepaspectratio = true]{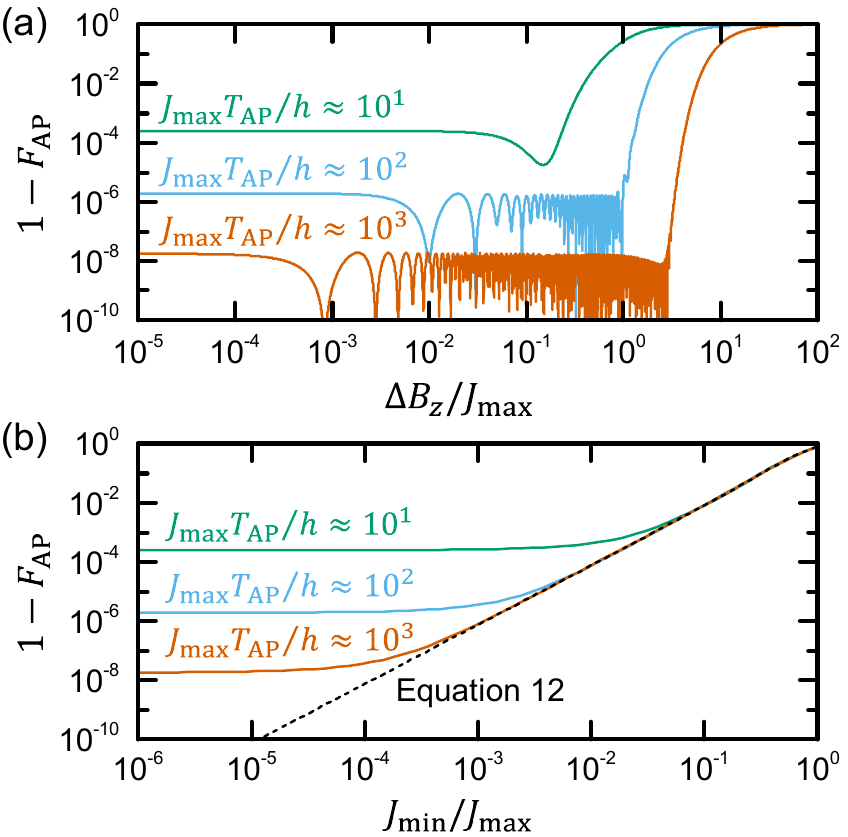}
\caption{
(Color online) The transport error due to (a) $\Delta B_z $ normalized to $J_\mathrm{max}$, and (b) limited tunability ($J_\mathrm{min}$/$J_\mathrm{max}$) of exchange couplings. These are obtained from numerical simulations of the adiabatic protocol for $J_\mathrm{max}T_{\mathrm{AP}}/h \approx$ 10$^1$, 10$^2$ and 10$^3$. The cut-off value of $\Delta B_z /J_\mathrm{max}$ such that $\Delta B_z$ does not affect the transport fidelity is given in Equation \ref{eq:AP_CutOff_DBz}. The dashed line in (b) is the error due to `spin-leakage' as described by Equation \ref{eq:Fidelity_AP_SpinLeakage}. }
\label{fig:APJminDBz}
\end{center}
\end{figure}

A non-zero $\Delta B_z$ decreases $\Delta E_{(-)}$ in Equation \ref{eq:APEnergySeparation}, and can therefore reduce the adiabaticity of the $\lvert{\downarrow_s}\rangle$ passage. Figure \ref{fig:APJminDBz}(a) plots the error for transporting a $\lvert\psi_s\rangle=\lvert\downarrow\rangle$ state, as a function of $\Delta B_z/J_\mathrm{max}$, for $J_\mathrm{max}T_{\mathrm{AP}}/h \approx$ 10$^1$, 10$^2$ and 10$^3$. The plot shows that for $\Delta B_z \ll J_\mathrm{max}$, the fidelities are limited only by $J_\mathrm{max}T_{\mathrm{AP}}/h$, and are given by Equation \ref{eq:Fidelity_AP_JmaxTAP}. As $\Delta B_z$ increases past a certain point, the reduction of $\Delta E_{(-)}$ causes the $\lvert \downarrow_s\rangle$ passage to lose adiabaticity. From Figure \ref{fig:APJminDBz}(a) we can extract the cut-off value for $\Delta B_z / J_\mathrm{max}$, such that $\Delta B_z$ does not reduce the transport fidelity, by fitting to the following expression:

\begin{equation}
\left(\frac{\Delta B_{z}}{J_\mathrm{max}}\right)_\mathrm{cut-off} \approx \frac{\sqrt{J_\mathrm{max}T_\mathrm{AP}/h}}{13}
\label{eq:AP_CutOff_DBz}
\end{equation}

Equation \ref{eq:AP_CutOff_DBz} shows that, that for $J_\mathrm{max}T_{\mathrm{AP}}/h \approx 10^1$, $\Delta B_z$ can be as large as $J_\mathrm{max}/4$ without affecting the fidelity. In the example of $\Delta B_z/h = 100$ MHz, $J_\mathrm{max}/h$ only needs to be as large as  $\approx$ 400 MHz. Even higher values of  $\Delta B_z/J_\mathrm{max}$ can be tolerated if $J_\mathrm{max}T_{\mathrm{AP}}/h$ is increased.

In addition to the adiabaticity, $\Delta B_z$ has another effect on the transported state. It breaks the symmetry of the $\lvert\uparrow\rangle$ and $\lvert\downarrow\rangle$ adiabatic passages, since $\Delta E_{(+)} \neq \Delta E_{(-)}$ in Equation \ref{eq:APEnergySeparation}. The transported state thus acquires a constant phase $\Delta\phi$ with respect to the initial source state. The phase $\Delta\phi$ can be calibrated and corrected for, because it is a function of $\Delta B_z$, $J_\mathrm{max}$, (which are determined during the calibration stage) and $T_\mathrm{AP}$.

\subsubsection{Limited tunability of exchange couplings}
\label{sec:AP_Jmin}

So far, we have assumed that the exchange coupling can be controlled up to the point of being entirely switched off, $J_\mathrm{min} = 0$. A more realistic assumption would allow for a limited dynamic range in the tunability of $J$, such that some residual exchange coupling remains at all times. For $J_\mathrm{min} > 0$, the eigenstates of the system at $t=0$ are not the simple tensor products of the source qubit state with the coupled states of the chain and target qubits. The transport fidelity then depends on how the system is initialized. If we assume that the system can be initialized such that the source qubit holds the state to be transported $\lvert \psi_s \rangle$ and the chain and target are in the `near singlet' state $\lvert\varphi_{S_{ct}}\rangle$, the initialized state is not an eigenstate. The source qubit will then undergo partial exchange oscillations (`spin-leakage') with the chain and target, resulting in an error.

To estimate the effect of this spin-leakage, we perform simulations where we wait for a time $h/J_\mathrm{max}$ after the transport protocol, which yields the worst-case fidelity. For simplicity, we assume $\Delta B_z = 0$. Figure \ref{fig:APJminDBz}(b) plots the error of transporting the $\lvert\psi_s\rangle = \lvert\uparrow\rangle$ state as a function of $J_\mathrm{min}/J_\mathrm{max}$, for $J_\mathrm{max}T_\mathrm{AP}/h \approx$ 10$^1$, 10$^2$ and 10$^3$. The fidelities calculated here are independent of the choice of $\lvert\psi_s\rangle$. We observe that the error traces in Figure \ref{fig:APJminDBz}(b) are first limited by Equation \ref{eq:Fidelity_AP_JmaxTAP} ($J_\mathrm{max}T_\mathrm{AP}/h$) and then by spin-leakage for large $J_\mathrm{min}/J_\mathrm{max}$. An analytical expression for the error due to spin-leakage is derived in Appendix \ref{appendix:fidelity_calculations}. This is plotted as the dashed black line in Figure \ref{fig:APJminDBz}(b) and is given by

\begin{equation}
1 - F_\mathrm{AP} \approx \frac{3}{4}\times(J_\mathrm{min}/J_\mathrm{max})^2.
\label{eq:Fidelity_AP_SpinLeakage}
\end{equation}

The quadratic dependence in Equation \ref{eq:Fidelity_AP_SpinLeakage} allows for low transport errors to be achieved with fairly limited tunability. One and two orders of magnitude of control over the exchange interaction result in errors of $10^{-2}$ and $\sim$ $10^{-4}$, respectively. Note that the exchange coupling that needs to be tuned in this system is that between a single donor electron operated in the bulk-like mode (source or target) and an electron at the edge of the chain operated in the interface mode. Estimation of the control on this exchange interaction is left for future work. A recent experiment has demonstrated limited tuning of the exchange coupling in a similar configuration, but the donor in that instance was located almost directly beneath the interface dot \cite{HarveyCollard2015arXiv}.

On another note, the spin-leakage error is essentially an estimate of the degree of isolation of the source qubit from the rest of the system. Figure \ref{fig:APJminDBz}(b) illustrates that good isolation can be achieved even with limited tunability of exchange couplings.  Recall the technique described in Section \ref{sec:Protocols} to isolate the source qubit during calibration, where $J_t$ is maximized with respect to $J_s$. With two orders of magnitude of tunability of exchange couplings, i.e. $J_t/J_s \sim 100$, the source qubit can be treated as being isolated with a fidelity of $\sim$ 99.99\%.

\begin{figure}[t!]
\begin{center}
\includegraphics[width=\columnwidth, keepaspectratio = true]{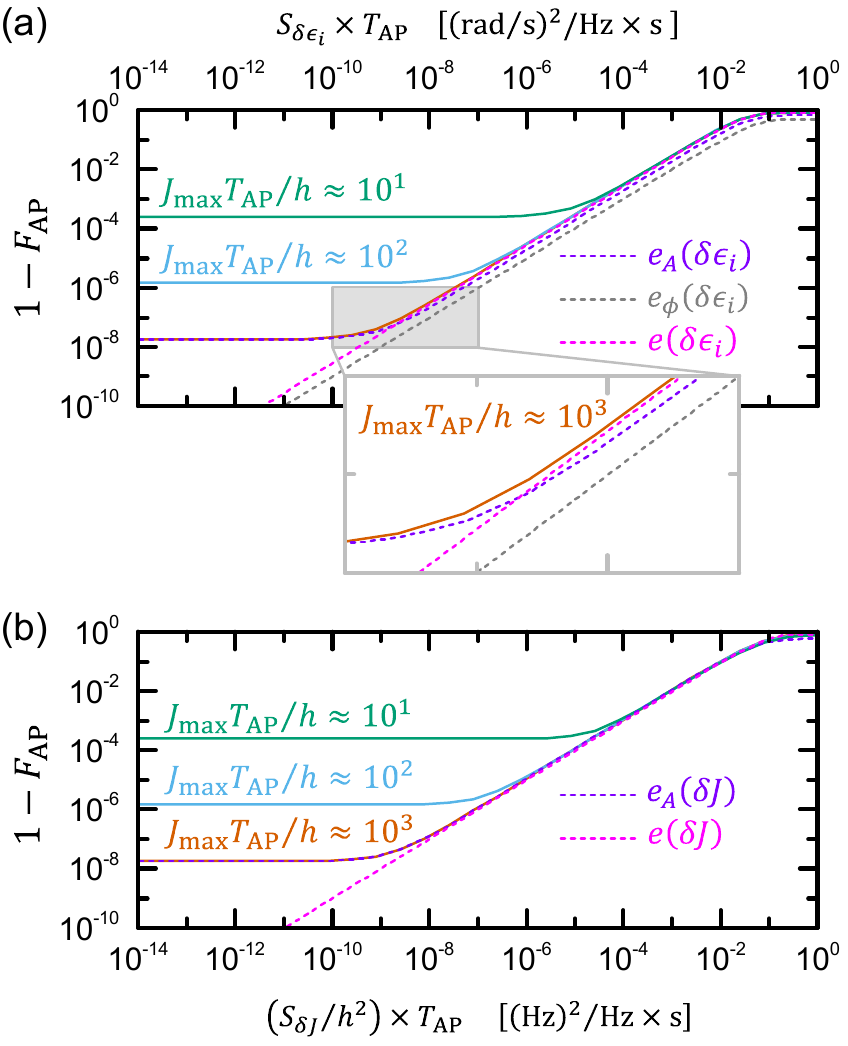}
\caption{(Color online) Calculations of error due to noise added to (a) $\epsilon_s$, $\epsilon_c$ and $\epsilon_t$ and (b) $J_s$ and $J_t$. The errors are plotted as a function of the product of the power spectral density of the white noise added ($S_{\delta\epsilon_i}$ and $S_{\delta J}$) and $T_\mathrm{AP}$. We assume $\Delta B_z/h = 100$ MHz. The solid lines are obtained from numerical simulations of the adiabatic protocol for $J_\mathrm{max}T_{\mathrm{AP}}/h \approx$ 10$^1$, 10$^2$ and 10$^3$. The dashed purple lines correspond to the error due to the loss of adiabaticity for $J_\mathrm{max}T_\mathrm{AP}/h\approx 10^3$. The dashed magenta lines are fits of the total error due to the noise given by Equations \ref{eq:Error_AP_noiseB0_fit} and \ref{eq:Error_AP_Jnoise_fit}. The dashed gray line in panel (a) shows the error due to dephasing.}
\label{fig:APnoise}
\end{center}
\end{figure}

\subsubsection{Noise on the qubit energy splittings $\epsilon_i$}

\label{sec:AP_noiseB0}

Magnetic and electric noise, arising both within the device and from the external control fields, results in fluctuations in $\epsilon_s$, $\epsilon_c$ and $\epsilon_t$. This can lower the fidelity of the transport protocol in two ways. First, the qubit being transported will be subject to dephasing, where an error in $\Delta\phi$ is accumulated. Second, higher-frequency noise can cause fast temporal variations in the instantaneous eigenstates, which can in turn make the transport passages lose adiabaticity.  

To estimate the effect of this noise, we simulate the adiabatic transport protocol with independent white Gaussian noise of power spectral density $S_{\delta\epsilon_i}$ added to $\epsilon_s$, $\epsilon_t$ and $\epsilon_c$ in Equation \ref{eq:Hamiltonian_source_chain_target}. For each value of $S_{\delta\epsilon_i}$, we perform 1000 simulations and compute the mean fidelity $F_\mathrm{AP}$ for transporting the $\lvert\psi_s\rangle = (\lvert\uparrow\rangle+\lvert\downarrow\rangle)/\sqrt{2}$ state, for $J_\mathrm{max}T_\mathrm{AP}/h \approx$ 10$^1$, 10$^2$ and 10$^3$, with $\Delta B_z/h = 100$ MHz. We plot these errors as a function of $\left(S_{\delta\epsilon_i} /h^2 \right) \times T_\mathrm{AP}$, as the solid lines in Figure \ref{fig:APnoise}(a). We express the noise added as $\left(S_{\delta\epsilon_i} /h^2 \right)$ to yield units of $\mathrm{Hz}^2/\mathrm{Hz}$, which is consistent with the quantities discussed in this paper.

We observe that, in the low-noise regime, the fidelities are limited by the value of $J_\mathrm{max}T_\mathrm{AP}/h$. As the level of noise increases, the fidelities are independent of $J_\mathrm{max}$ and instead only depend on the magnitude of the noise and the time that the noise has to act on the system.

The total error is a combination of errors due to dephasing and loss of adiabaticity. To capture the error due to the loss of adiabaticity alone, we perform a separate simulation to obtain the average error $e_A(\delta\epsilon_i)$ of transporting the $\lvert\psi_s\rangle = \lvert\uparrow\rangle$ and $\lvert\psi_s\rangle = \lvert\downarrow\rangle$ states, which are immune to dephasing. $e_A(\delta\epsilon_i)$ is computed for $J_\mathrm{max}T_\mathrm{AP}/h \approx$ 10$^3$ and is plotted as the dashed purple line in Figure \ref{fig:APnoise}(a). We observe that this matches the solid orange line when $\left(S_{\delta\epsilon_i} /h^2 \right) \times T_\mathrm{AP} < 10^{-10}$. Beyond this value, dephasing also contributes to the total error, separating these two lines, as shown by the inset.

The error due to dephasing of a single qubit,  $e_\phi(\delta\epsilon_i)$ in the presence of white noise is given by \cite{Yuge2011prb}

\begin{equation}
\begin{split}
e_\phi(\delta\epsilon_i) = \frac{ 1 - e^{-T_\mathrm{AP}/T_2}}{2} = \frac{1 - e^{-2\pi^2S_{\delta\epsilon_i}T_\mathrm{AP}/h^2}}{2},
\end{split}
\label{eq:Error_AP_noiseB0_dephasing}
\end{equation}

where $T_2$ is the qubit dephasing time. We plot $e_\phi(\delta\epsilon_i)$ as the dashed gray line in Figure \ref{fig:APnoise}(a). Assuming the two sources of error to be independent, the total error is given by $e_{\phi}(\delta\epsilon_i) + e_{A}(\delta\epsilon_i) - e_{\phi}(\delta\epsilon_i)e_{A}(\delta\epsilon_i)$. This matches the solid orange line remarkably well (result not plotted). In the regime where the transport is limited by noise, we fit the error to an exponential function. This yields the fit, $e(\delta\epsilon_i)$ (plotted as the dashed magenta line in Figure \ref{fig:APnoise}(a)), given by

\begin{equation}
e(\delta\epsilon_i) \approx 0.83\left(1 - e^{-34\times S_{\delta\epsilon_i}T_\mathrm{AP}/h^2}\right).
\label{eq:Error_AP_noiseB0_fit}
\end{equation}

Comparing these results to the experiment requires knowledge of the frequency-dependence of the power spectral density of the noise. A recent experiment reports the noise spectrum for a donor electron spin qubit in isotopically enriched $^{28}$Si to be of the form $9\times 10^{11}/\omega^{2.5} + 6$ (rad/s)$^2$/Hz. The frequency-dependent component is attributed to fluctuations in the external magnetic field $B_0$ \cite{Muhonen2014nn}, which would be homogeneous over the typical transport length-scales ($\sim100$ nm). This effect of this noise component can therefore be refocused with dynamical decoupling \cite{Viola1999prl}. On the contrary, white noise cannot be refocused. With the reported noise floor of 6 (rad/s)$^2$/Hz, such that $\left(S_{\delta\epsilon_i}/h^2\right) \approx 0.15$, Equation \ref{eq:Error_AP_noiseB0_fit} predicts errors of $\sim 10^{-7}$ with $T_\mathrm{AP} \approx 100$~ns.

\subsubsection{Noise in the qubit-chain couplings $J_s$ and $J_t$}

\label{sec:AP_noiseJ}

Electrical noise in a gated nanostructure can modify the exchange interactions $J_s$ and $J_t$. For $\Delta B_z = 0$, noise in $J_s$ and $J_t$ modifies the $\lvert\uparrow\rangle$ and $\lvert\downarrow\rangle$ adiabatic passages equally, such that the phase error of the transported state is zero. However, $\Delta B_z/h \sim 100$ MHz is always finite for an interface-mode chain, breaking the symmetry of the passages and allowing this noise to potentially feed in to the phase of the transported qubit. Additionally, high frequency noise can potentially reduce the adiabaticity of transport by rapidly modifying the instantaneous eigenstates.

We estimate the transport fidelity as a function of $\left(S_{\delta J} /h^2 \right) \times T_\mathrm{AP}$, where $S_{\delta J}$ is the power spectral density of white noise added to $J_s$ and $J_t$. For each value of $S_{\delta J}$, we perform a Monte Carlo analysis of 1000 simulations to quantify the fidelity $F_\mathrm{AP}$ of transporting the $\lvert\psi_s\rangle=\left(\lvert{\downarrow}\rangle + \lvert{\uparrow}\rangle\right)/\sqrt{2}$ state with $\Delta B_z/h = 100$ MHz. These errors are plotted as the solid lines in Figure \ref{fig:APnoise}(b), for $J_\mathrm{max}T_\mathrm{AP}/h \approx $ 10$^1$, 10$^2$ and 10$^3$. The trend observed is the same as that of Figure \ref{fig:APnoise}(a). Note that solid orange line is hidden by the dashed purple line.

To quantify the loss of adiabaticity, we obtain the average error $e_A(\delta J)$ of transporting the $\lvert\psi_s\rangle = \lvert\uparrow\rangle$ and $\lvert\psi_s\rangle = \lvert\downarrow\rangle$ states, as they are immune to dephasing. $e_A(\delta J)$ for $J_\mathrm{max}T_\mathrm{AP}/h \approx$ 10$^3$ is plotted as the dashed purple line in Figure \ref{fig:APnoise}(b). It aligns almost exactly with the solid orange line for $\left(S_{\delta J} /h^2 \right) \times T_\mathrm{AP} < 10^{-2}$, indicating that the loss of adiabaticity is the main source of error for noise in $J_s$ and $J_t$. In the regime where the transport is limited by noise, we fit the error to an exponential function. This yields the fit, $e(\delta J)$ (plotted as the dashed magenta line in Figure \ref{fig:APnoise}(b), given by

\begin{equation}
e(\delta J) \approx 0.83\left(1 - e^{-11S_{\delta J}T_\mathrm{AP}/h^2}\right).
\label{eq:Error_AP_Jnoise_fit}
\end{equation}

Experimental values for noise on the exchange coupling between donors are not currently available. In any case, our analysis shows that it is favorable to perform the transport protocol in shorter times $T_\mathrm{AP}$ and larger $J_\mathrm{max}$.\\

Overall, we find that high-fidelity spin-qubit transport across donor chains may be achieved with the adiabatic protocol. This protocol is inherently robust to errors in the precise magnitudes of exchange couplings and the transport time (Figure \ref{fig:APFidelity_JmaxTAP}). The inclusion of $\Delta B_z$ in the system is utilized to initialize the system for transport. For $\Delta B_z/h \sim 100$ MHz, we find that a minimum $J_\mathrm{max}/h$ of 400 MHz is sufficient to ensure that the fidelity is unaffected by $\Delta B_z$ (Figure \ref{fig:APJminDBz}(a)). In the case of limited tunability of the exchange coupling, we have found that two orders of magnitude of control is sufficient for fidelities exceeding 99.99\% (Figure \ref{fig:APJminDBz}(b)). The magnitude of magnetic noise as measured in a recent experiment in isotopically-purified silicon still allows for errors $\sim 10^{-7}$ to be achieved (Figure \ref{fig:APnoise}(a)). As for the noise in the exchange couplings, although we have calculated the transport fidelities as a function of the noise power spectral density, we do not have compatible experimental measurements for comparison.

\section{SUMMARY AND OUTLOOK}

We have provided a comprehensive analysis of the operation of an odd-number donor spin chain for the purpose of transporting a spin qubit state across a quantum processor. A key realization is that, while the donor placement accuracy necessary to operate a spin chain in the bulk-like mode imposes extremely tight constraints on donor placement, a much more reliable fabrication pathway can be found by adopting the interface-mode operation. In that mode, the donor placement accuracy achievable with ion implantation process can allow the fabrication of functional spin chains with high yield. Moreover, because of the wider extent of the electron wave function at the interface, the qubit state can be moved across distance of order 100~nm using a modest number of donors. Because of the absence of hyperfine coupling between donor nuclear spins and their respective electrons while confined at the Si-SiO$_2$ interface, the system has an inbuilt difference in energy splitting between the source/target qubits and the chain that links them. We have shown how to use this property to initialize the system in a state useful for adiabatic transport of a qubit spin state.

Our analysis of the realistic noise sources that could be present in a spin chain device, based upon the existing knowledge of such noise sources in donor spin qubit devices, indicates that spin transport with high fidelity is in principle possible. Therefore, future work can focus on the design and development of large-scale quantum computer architectures where highly coherent donor spin qubits are linked by spin chains. In that context, the method discussed in Section \ref{sec:AP_Jmin} to isolate individual qubits from their neighbors may become more broadly significant, because controlling and removing unwanted interactions between physical qubits is vital to the high-fidelity operation of a quantum computer. For example, a combination of single donors and donor chains could be used to isolate information-carrying spins when they are required to be idle. The adiabatic protocol can then be used within the same system to transport these spins to appropriate locations, where they interact with other qubits to perform quantum logic operations.

\begin{acknowledgements}
This research was funded by the Australian Research Council Centre of Excellence for Quantum Computation and Communication Technology (project number CE11E0001027) and the US Army Research Office under contract number W911NF-13-1-0024. NCN/nanohub.org computational resources funded by the National Science Foundation under contract number EEC-0634750 were used in this work. F. A. M. and R. K. contributed equally to this work.
\end{acknowledgements}

\appendix
\section{Exchange coupling calculation with NEMO-3D}
\label{appendix:exchange_calculations}
In Section \ref{sec:operation_donor_chain} of the manuscript, we estimated the required dopant placement accuracy to successfully realize a donor chain. Part of this calculation involved a numerical estimate of the exchange coupling $J$ as a function of donor separation.  For this, we consider two donors A and B placed in a 3-dimensional space $\mathbf{r}$. We first calculate the single-electron wavefunctions $\varPsi_{A}(\mathbf{r})$ and $\varPsi_{B}(\mathbf{r})$ independently for the two donors using NEMO-3D -- an  atomistic tight-binding simulation package \cite{Klimeck2007ieeeteda, Klimeck2007ieeetedb}. To estimate $J$ between the two electrons, we use the Heitler-London formula: \cite{slater1965quantum, wellard2005prb,  Rahman_phd_thesis}

\begin{subequations}
\begin{equation}
J = \frac{2}{1 - |S_0|^4}\{|S_0|^2J_0 - K_0\},
\end{equation}
\begin{equation}
S_0 = \int\limits_{\mathbf{r}}\varPsi_{A}(\mathbf{r})^*\varPsi_{B}(\mathbf{r})~d\mathbf{r}
\end{equation}
\begin{equation}
\begin{split}
J_0 = \int\limits_{\mathbf{r_1}} \int\limits_{\mathbf{r_2}}\varPsi_{A}(\mathbf{r_1})^*\varPsi_{A}(\mathbf{r_1})\frac{q^2}{4\pi \epsilon_\mathrm{Si}|\mathbf{r_1} - \mathbf{r_2}|}\\\varPsi_{B}^*(\mathbf{r_2})\varPsi_{B}(\mathbf{r_2})~d\mathbf{r_2}d\mathbf{r_1},
\end{split}
\end{equation}
\begin{equation}
\begin{split}
K_0 = \int\limits_{\mathbf{r_1}}\int\limits_{\mathbf{r_2}}\varPsi_{A}(\mathbf{r_1})^*\varPsi_{B}(\mathbf{r_1}) \frac{q^2}{4\pi \epsilon_\mathrm{Si}|\mathbf{r_1} - \mathbf{r_2}|}\\\varPsi_{B}^*(\mathbf{r_2})\varPsi_{A}(\mathbf{r_2})~d\mathbf{r_2}d\mathbf{r_1},
\end{split}
\end{equation}
\label{eq:HL_J_expression}
\end{subequations}

where $J_0$ and $K_0$ are commonly referred to as the exchange and Coulomb integrals, respectively. $q$ is the charge of the electron, and $\epsilon_\mathrm{Si}$ is the dielectric constant of silicon. Note that the wavefunctions $\varPsi_{A}(\mathbf{r})$ and $\varPsi_{B}(\mathbf{r})$ in this method are computed independently of each other. This is reasonable provided the separation between the donors is several times the Bohr radii.

\begin{figure}[t!]
\begin{center}
\includegraphics[width=\columnwidth, keepaspectratio = true]{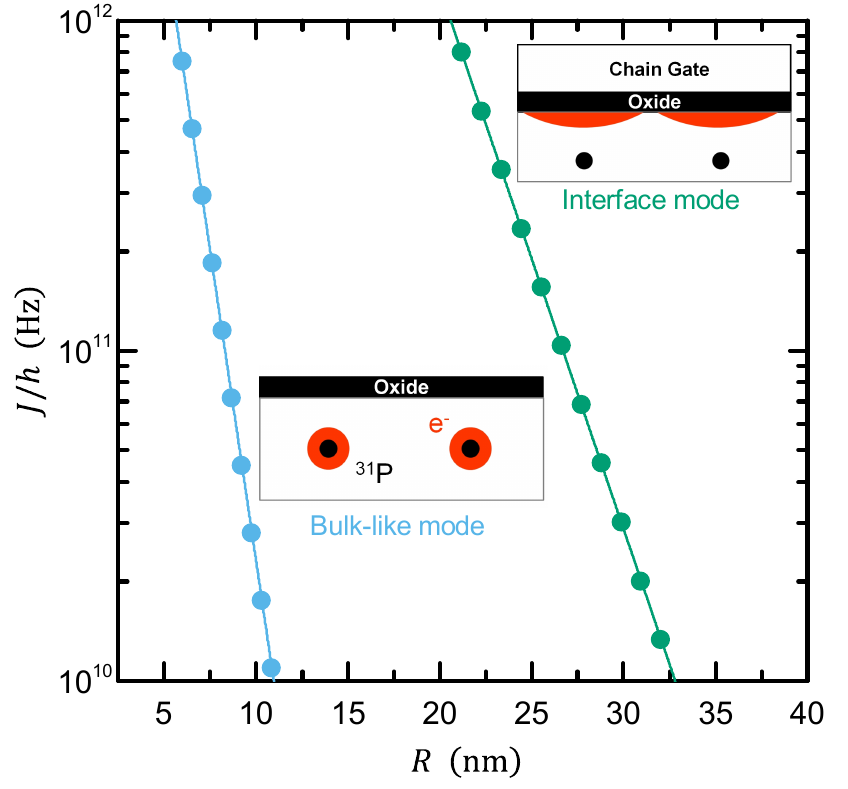}
\caption{(Color online) Tight-binding calculation of the exchange coupling as a function of donor separation, for the bulk-like (blue) and interface (green) modes of operation modes. Insets: Schematics of the electron orbital wavefunctions for the two operation modes, with the electrons and $\mathrm{^{31}P}$ nuclei represented in red and black, respectively. }
\label{fig:NEMO_exchange}
\end{center}
\end{figure}

Figure \ref{fig:NEMO_exchange} plots $J$ as a function of donor separation $R$, assuming the donors to be placed along the [100] plane at a depth 7.1 nm below a Si-SiO$_2$ interface. Here, we focus on the regime where 10 GHz $<J/h<$ 1 THz, and consider the two operating modes: (i) the bulk-like mode where the electrons are bound to their respective donors, and (ii) the interface mode where the electrons are pulled towards the Si-SiO$_2$ interface with a vertical electric field $E_z = 30$ MV/m. As the donor electrons are pulled to the interface in the interface mode, their wavefunctions expand in the lateral direction (illustrated in the insets of Figure \ref{fig:NEMO_exchange}), causing $J$ to be enhanced by many orders of magnitude. The interface-mode therefore allows the donors to be separated further apart, while maintaining large exchange couplings within the chain. The dependence of $J$ on donor separation $R$ in Figure \ref{fig:NEMO_exchange} can be fitted to an exponential function given by

\begin{equation}
J = J_{\lambda}e^{-R/R_{\lambda}}
\label{eq:tight_binding_exchange}
\end{equation}

In the bulk-like mode, $J_{\lambda}/h = 119.12$  THz and $R_\lambda = 1.17$ nm. In the interface mode, $J_{\lambda}/h =$ 2.34 PHz and $R_{\lambda} =$ 2.64 nm.

\section{Donor chain operated in the bulk-like mode}
\label{appendix:bulk_like_mode}

The nuclear spins of the chain donors do not influence the qubit transport when the chain is operated in the interface mode. However, they have an important effect for a chain operated in the bulk-like mode, since they affect the qubit energy splitting $\epsilon_c$ in Equation \ref{eq:Hamiltonian_source_chain_target}. Here we describe the dependence of $\epsilon_c$ on the state of the chain nuclei.

The Hamiltonian for a chain consisting of $N$ donors, including nuclear spins, is given by

\begin{equation}
H_c = H_{c_e} - \sum\limits_{i=1}^{N}  h\gamma_n B_0 \frac{^\mathrm{nuc}\sigma_{z,c(i)}}{2} + \sum\limits_{i=1}^{N}  A_{c(i)} \frac{\mathrm{\boldsymbol{\sigma}}_{c(i)}}{2}    \cdot    \frac{^\mathrm{nuc}\mathrm{\boldsymbol{\sigma}}_{c(i)}}{2}
\label{eq:Hamiltonian_chain_bulk}
\end{equation}

where $H_{c_e}$ is given by Equation \ref{eq:electron_chain_Hamiltonian}, $^\mathrm{nuc}\mathrm{\boldsymbol{\sigma}}_{c(i)}$ is the Pauli operator for the $i^\mathrm{th}$ chain nucleus with $z$-component $^\mathrm{nuc}\sigma_{z,c(i)}$. $A_{c(i)}$ is the hyperfine coupling between the electron and nuclear spins of the $i^{\mathrm{th}}$ chain donor. Recall that for a large magnetic field, e.g. $B_0 = 1$ T, the electron and nuclear spin states can be treated separately. The $N$ electrons form the extended qubit described in Section \ref{sec:operation_donor_chain}, provided Equation \ref{eq:chain_criterion} is satisfied. For each state of the chain qubit, there are therefore $2^N$ eigenstates for the nuclei. The resulting ESR spectrum of the chain consists of $2^N$ resonances, where the chain qubit state is flipped conditional on the state of the nuclear system.

\begin{figure}[t!]
\begin{center}
\includegraphics[width=\columnwidth, keepaspectratio = true]{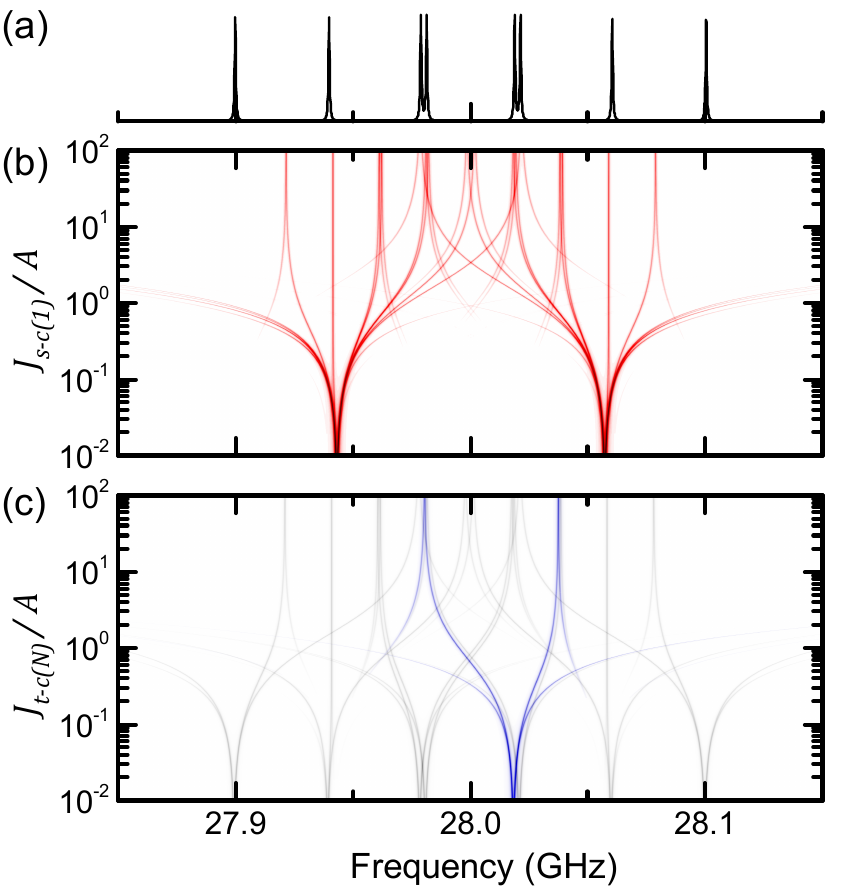}
\caption{(Color online (a) Electron spin resonance (ESR) frequencies for a three-donor chain operated in the bulk-like mode, with $B_0 = 1$ T and $J_{c}/h = 1$ THz. (b) ESR spectrum of a donor coupled to the edge of a $N=3$ donor chain. (c) ESR spectrum of a $N=3$ chain qubit coupled to a single donor, i.e. assuming that the chain, not the donor, is being measured.}
\label{fig:ESRspectrum_source_chain_bulk}
\end{center}
\end{figure}

For example, we plot the ESR spectrum for the example of a 3-donor spin chain operated in the bulk-like mode with $B_0=1$ T in Figure ~\ref{fig:ESRspectrum_source_chain_bulk}(a). The eight ESR transitions reveal the hyperfine shifts in the chain qubit resonance frequency from $\gamma_e B_0 = 28$ GHz. The frequency shift, which we denote as $\Delta\nu_c$, is a function of the nuclear state and the individual hyperfine couplings.

We can provide an expression for this shift by first determining the eigenstates of the nuclei. The chain nuclei are mutually coupled by an electron-mediated super-exchange coupling, $J_n$, which is a function of the individual hyperfine couplings $A_{c(i)}$ and the electronic exchange couplings $J_{c(i)}$ \cite{Kane1998n}. For a 3-donor chain with $J_{c(i)}/h = 1$~THz and $A_{c(i)}/h = 100$ MHz, we numerically calculate that $J_n/h$ between the first and third nuclei is $\sim 100$ kHz. However, in realistic devices, local electric fields and strain can introduce a Stark shift of order a few MHz in the individual hyperfine couplings $A_{c(i)}/h$  \cite{Mohiyaddin2013nl, laucht2015sciadv}, which in turn detune the nuclei from each other by an amount that typically exceeds the magnitude of their mutual couplings. We account for this by introducing variations in the hyperfine couplings of order 1 MHz in Equation \ref{eq:Hamiltonian_chain_bulk}. The $\sim$MHz detuning dominates over the weak coupling $J_n/h$, such that the nuclear eigenstates are almost exactly the tensor products of their individual $\lvert\Uparrow\rangle$ and $\lvert\Downarrow\rangle$ states. We can thus calculate $\Delta\nu_c$ to first order using the equation below.

\begin{equation}
\Delta\nu_c =  \sum\limits_{i=1}^{N} U(i) A_{c(i)}C_{c(i)}/h,
\label{eq:Chain_bulk_gamman}
\end{equation}

where $U(i)$ equals 1 or -1 when the $i^\mathrm{th}$ nuclear spin is $\lvert\Uparrow\rangle$ or $\lvert\Downarrow\rangle$, respectively. $2C_{c(i)} = \langle\uparrow_c\rvert \sigma_{zc(i)}\lvert\uparrow_c\rangle = -\langle\downarrow_c \rvert \sigma_{zc(i)} \lvert \downarrow_c\rangle$ represents the `effective contribution' of the $i^{\mathrm{th}}$ chain electron to the chain spin-1/2 ground states. With this result, we can map the effect of the nuclei for a chain operated in the bulk-like mode onto a shift in $\epsilon_c$ by an amount $h\Delta\nu_c$.

\section{Calibration of the donor chain in the bulk-like mode}
\label{appendix:bulk_like_mode_calibration}

In Section \ref{sec:Protocols}, we assumed that the chain is operated in the interface mode while calibrating the exchange couplings and the individual qubit energy splittings. This was because experimentally realizing a chain operated in the bulk-like mode is extremely challenging compared to realizing one operated in the interface mode. Here, for completeness, we outline the calibration protocols when the chain is operated in the bulk-like mode.

The architecture including the SET, source donor, donor chain and target donors is identical to that described in Section \ref{sec:Protocols} of the main text. The difference between the interface and bulk-like modes first appears when measuring the exchange coupling $J_s$ between the source and chain qubits. The ESR spectrum of the source electron will not only show contributions from the state of the chain qubit, but also from the state of the chain nuclei. Figure \ref{fig:ESRspectrum_source_chain_bulk}(b) shows the ESR spectrum for a source donor coupled to a 3-donor chain, as a function of the exchange coupling $J_{s-c(1)}$ between the source and the first element of the chain. This spectrum is calculated by numerically solving for the eigenstates of the following Hamiltonian.

\begin{equation}
\begin{split}
H_{s-c(\mathrm{bulk})} = H_c + h\gamma_eB_0 \frac{\sigma_{z,s}}{2} - h\gamma_nB_0 \frac{^\mathrm{nuc}\sigma_{z,s}}{2} + \\ A_s \frac{\mathrm{\boldsymbol{\sigma}}_{s}}{2}    \cdot    \frac{^\mathrm{nuc}\mathrm{\boldsymbol{\sigma}}_{s}}{2} + J_{s-c(1)}    \frac{\mathrm{\boldsymbol{\sigma}}_{s}}{2}    \cdot    \frac{\mathrm{\boldsymbol{\sigma}}_{c(1)}}{2},
\end{split}
\label{eq:Hamiltonian_source_interface_bulk}
\end{equation}

Note that $J_s \propto J_{s-c(1)}$ as described in Section \ref{sec:operation_donor_chain}. For each value of $J_{s-c(1)}$, we find all allowed transitions between eigenstates of $H_{s-c(\mathrm{bulk})}$  and weigh them with the product of the transition probability and the spin readout contrast of the source qubit, as done for the interface mode.

Experimentally, to observe all transition frequencies in the ESR spectrum, the chain qubit and nuclei need to be randomized. The NMR frequencies of an $N$-donor chain are bound between $0$ and $\gamma_n B_0 + A/2h$, as will be explained in Appendix \ref{appendix:chain_NMR_spectrum}. This makes it possible to randomize the nuclei with non-adiabatic sweeps over the NMR frequencies. Similarly, non-adiabatic sweeps over the ESR frequencies in Figure \ref{fig:ESRspectrum_source_chain_bulk}(a) can be used to randomize the chain qubit.

Figure \ref{fig:ESRspectrum_source_chain_bulk}(b) shows that in the low $J_{s-c(1)}$ regime, the ESR spectrum of the source electron consists of two transition frequencies corresponding to the two states of its nuclear spin. As $J_{s-c(1)}$ is increased slightly, these split into a pair of branches due to the coupling to the chain qubit, which can be in either the $\lvert \uparrow_c \rangle$ or $\lvert \downarrow_c \rangle$ state. The branches are split by $J_s/h$, allowing $J_s$ to be calibrated from the ESR spectrum. For large $J_{s-c(1)}$ ($\gg A$), the branches involving the $\lvert T_0 \rangle$-like state tend towards the average of the uncoupled source and chain qubit frequencies,  $(\epsilon_s + \epsilon_c)/2h$. This results in $2 \times 2^N$ possible frequencies, since there are $N$ nuclei in the chain. For $J_s \gg A$, $J_s$ can be calibrated using the SWAP-style experiments outlined in Section \ref{sec:Protocols}. Once $J_s$ has been calibrated, $\epsilon_c$ (and therefore $\Delta B_z$) can be determined by performing ESR on the isolated chain qubit, as explained in Section \ref{sec:Protocols}.

Figure ~\ref{fig:ESRspectrum_source_chain_bulk}(c) then shows the ESR spectrum of the chain qubit coupled to a target donor, for the purpose of calibrating $J_t$. In the low $J_t$ regime, the chain qubit has eight possible transition frequencies corresponding to those in Figure ~\ref{fig:ESRspectrum_source_chain_bulk}(a). Highlighted in bright blue are the branches corresponding to a particular nuclear configuration ($\lvert\Downarrow\Downarrow\Uparrow\rangle$) of the chain. This was done to simplify the understanding of the spectrum, as well as to show that $J_t$ can be obtained even if the chain nuclei are not randomized.

\section{NMR frequencies of a donor chain in the bulk-like mode}
\label{appendix:chain_NMR_spectrum}
Recall from Appendix \ref{appendix:bulk_like_mode} that the nuclear eigenstates of a realistic donor chain in bulk-like mode would be tensor products of $\lvert\Uparrow\rangle$ and $\lvert\Downarrow\rangle$ states of the chain nuclei. The NMR frequency $\nu_{ni}$ for flipping the $i^\mathrm{th}$ chain nucleus to first order is given by

\begin{equation}
\nu_{ni} = \lvert -\gamma_n B_0 \pm C_{c(i)} A/h \rvert ,
\label{eq:chain_NMR_frequency}
\end{equation}

where $C_{c(i)}$ is as defined in Equation \ref{eq:Chain_bulk_gamman}. The sign is $+$ or $-$ when the chain qubit state is $\lvert\uparrow_c\rangle$ or $\lvert\downarrow_c\rangle$, respectively.

$C_{c(i)}$ takes its maximum value of 0.5 when $N = 1$. Hence the maximum value of $\nu_{ni}$ is $(\gamma_n B_0 + A/2h)$ for any $N$. Thus randomized non-adiabatic NMR frequency sweeps from DC to past this maximum value would be sufficient to randomize the nuclei for a donor chain of any size. Practically, the sweep involves applying a frequency-modulated excitation where the rate of change of frequency is faster than, but comparable to, the expected Rabi frequency of a nuclear spin \cite{Laucht2014apl}.

\section{Fidelity Calculations}
\label{appendix:fidelity_calculations}

In Sections \ref{sec:SingletInitialization} and \ref{sec:adiabatic_transport}, we quantified the fidelities of the singlet initialization protocol and the adiabatic transport protocol, respectively. Here, we outline the derivations used to obtain Equations \ref{eq:Fidelity_SP}, \ref{eq:Fidelity_AP_JmaxTAP} and \ref{eq:Fidelity_AP_SpinLeakage}.

\subsection{Singlet Initialization Fidelity}
\label{appendix:SPfid}

\begin{figure}[t!]
	\begin{center}
		\includegraphics[width=\columnwidth, keepaspectratio = true]{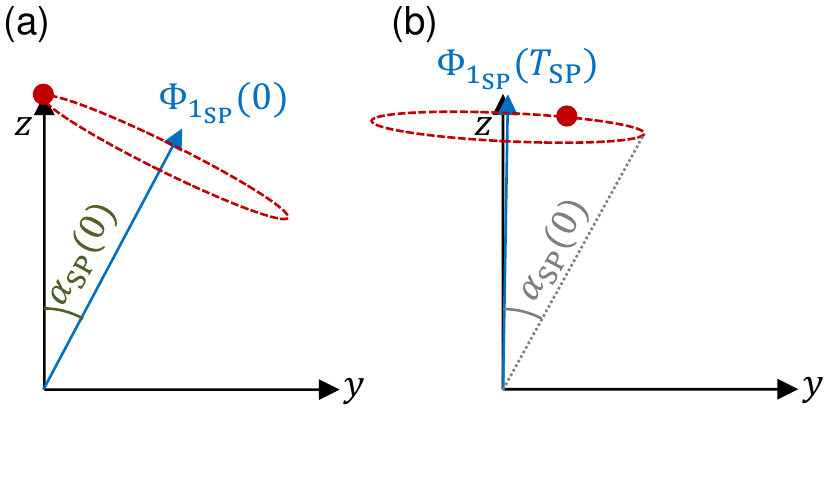}
		\caption{(Color online) Bloch sphere in the adiabatic frame at the (a) start and (b) end of the singlet initialization protocol. The red dot represents the instantaneous state. The initial state of the system is oriented along the $z$-axis. In the adiabatic limit, the `circle of precession' (dashed red circle) around the eigenstate $\lvert\Phi_{1_\mathrm{SP}}(t)\rangle$ (blue arrow) follows the evolution of $\lvert\Phi_{1_\mathrm{SP}}(t)\rangle$ and does not change in diameter.}
		\label{fig:SPcircleOfPrecession}
	\end{center}
\end{figure}

The singlet initialization protocol involves initializing the chain-target system in the $\lvert\widetilde{\downarrow_c\uparrow_t}\rangle$ state, as defined in Section \ref{sec:SingletInitialization}, and ramping the exchange coupling between the chain and target qubits. The $\lvert\uparrow_c\uparrow_t\rangle$ and $\lvert\downarrow_c\downarrow_t\rangle$ states are not coupled to either the $\lvert\uparrow_c\downarrow_t\rangle$ or $\lvert\downarrow_c\uparrow_t\rangle$ states. Therefore, the dynamics of the protocol can be represented in the $S$-$T_0$ Bloch sphere, as illustrated in Figure \ref{fig:APSingletInitialization}(a). The truncated Hamiltonian in the basis $\{ \lvert T_{0ct}\rangle, \lvert S_{ct}\rangle  \}$ is given by

\begin{equation}
H_\mathrm{SP}(t) = J(t)\frac{\sigma_z}{2} + \Delta B_z\frac{\sigma_x}{2},
\label{eq:Singlet_Initialization_Simplified_Hamiltonian}
\end{equation}

where $J(t)$ is the exchange coupling between the chain and target, and is linearly ramped from 0 to $J_\mathrm{max}$. The aim is to have the system in the $\lvert\varphi_{S_{ct}}\rangle$ eigenstate at the end of the protocol, as defined in Section \ref{sec:SingletInitialization}. Therefore, to quantify the fidelity of the protocol, we translate the Hamiltonian into the adiabatic frame. The translated Hamiltonian $H_\mathrm{SP}^A(t)$ is obtained by applying the following operation

\begin{equation}
H_\mathrm{SP}^A(t) = AH_\mathrm{SP}A^{-1} - i\hbar A\frac{d}{dt}(A^{-1}),
\label{eq:Singlet_Initialization_adiabatic_frame}
\end{equation}

where the row vectors of $A$ are the time-varying eigenvectors of $H_\mathrm{SP}(t)$.

\begin{equation}
H_\mathrm{SP}^A(t) = \sqrt{J(t)^2 + \left(\Delta B_z\right)^2}\frac{\sigma_z}{2} - \hbar\frac{d\chi_\mathrm{SP}}{dt}\frac{\sigma_y}{2},
\label{eq:Singlet_Initialization_adiabatic_frame_simplified}
\end{equation}

where $\mathrm{tan}(\chi_\mathrm{SP}) = \Delta B_z/J(t)$. To gain insight into the energy terms in $H_\mathrm{SP}^A$, it is instructive to compare the system in the adiabatic frame onto a spin in a magnetic field (Figure \ref{fig:SPcircleOfPrecession}). In this picture, the field in the $z$-direction is the energy separation of the eigenstates of $H_\mathrm{SP}$. The field in the $y$-direction is proportional to the rate of change of the angle of the eigenstate in the Bloch sphere of $H_\mathrm{SP}$ (the laboratory frame). The eigenvectors of $H_\mathrm{SP}^A(t)$ are given by

\begin{subequations}
\begin{equation}
\lvert\Phi_{1_\mathrm{SP}}(t)\rangle =
  \begin{pmatrix}
    \mathrm{cos}(\alpha_\mathrm{SP})\\
    -i\mathrm{sin}(\alpha_\mathrm{SP})
  \end{pmatrix},
\end{equation}
\begin{equation}
\lvert\Phi_{2_\mathrm{SP}}(t)\rangle =
  \begin{pmatrix}
    \mathrm{sin}(\alpha_\mathrm{SP})\\
    i\mathrm{cos}(\alpha_\mathrm{SP})
  \end{pmatrix},
\end{equation}
\label{eq:Singlet_Initialization_adiabatic_frame_eigen_state}
\end{subequations}

where $\mathrm{tan}(2\alpha_\mathrm{SP}) = \hbar\frac{d\chi_\mathrm{SP}}{dt}/\sqrt{J(t)^2 + \Delta B_z^2}$. The fidelity is determined by the closeness of the state at the end of the protocol to the $+z$-axis in the adiabatic frame, which is equivalent to the eigenstate of $H_\mathrm{SP}$.

The dynamics of the system in the adiabatic frame, $H_\mathrm{SP}^A$, is shown in Figure \ref{fig:SPcircleOfPrecession}. Figure \ref{fig:SPcircleOfPrecession}(a) shows the initial state of the system, which is oriented along $z$, as we start in an eigenstate of $H_\mathrm{SP}$. However, the field along $y$ is non-zero at this point, such that the eigenstate $\lvert\Phi_{1_\mathrm{SP}}(t)\rangle$ is at an angle $\alpha_\mathrm{SP}(0)$ from the $z$-axis. If we consider the protocol to be in the adiabatic limit, then the precession frequency of the initial state around the eigenstate $\lvert\Phi_{1_\mathrm{SP}}(t)\rangle$ is much faster than $d\alpha_\mathrm{SP}(t)/dt$. Hence, we can picture the precession trajectory of the state to be a circle around  $\lvert\Phi_{1_\mathrm{SP}}(t)\rangle$ (dashed red circle in Figure \ref{fig:SPcircleOfPrecession}). In the adiabatic limit, the center of this `circle of precession' follows $\lvert\Phi_{1_\mathrm{SP}}(t)\rangle$, and therefore the projection of the instantaneous state onto $\lvert\Phi_{1_\mathrm{SP}}(t)\rangle$ remains constant.

From Equation \ref{eq:Singlet_Initialization_adiabatic_frame_eigen_state}, we see that $\alpha_\mathrm{SP}$ tends towards zero at $t=T_\mathrm{SP}$, as also illustrated in Figure \ref{fig:SPcircleOfPrecession}(b). At the end of the protocol, the circle of precession is centered close to the $z$-axis, with its diameter set by $\alpha_\mathrm{SP}(0)$. Therefore, the squared-projection of the eigenstate at $t=0$ on to the $z$-axis is an estimate of the fidelity of the protocol, and is given by

\begin{equation}
1 - F_\mathrm{SP} \approx \mathrm{sin}^2(\alpha_\mathrm{SP}(0)) = \frac{1}{2} {\left({1 - \sqrt{\frac{K}{1 + K}}}\right)},
\end{equation}

where $K =\left(\Delta B_z/J_\mathrm{max}\right)^4\times\left( J_\mathrm{max}T_\mathrm{SP} /\hbar \right)^2$. In the limit where $K \gg 1$, this can be simplified to

\begin{equation}
1 - F_\mathrm{SP} \approx \frac{1}{4\left( J_\mathrm{max}T_\mathrm{SP}/\hbar\right)^2} \left(\frac{J_\mathrm{max}}{\Delta B_z}\right)^4.
\end{equation}

\subsection{Adiabatic Transport Fidelity}

\subsubsection{Adiabaticity errors: $J_\mathrm{max}T_\mathrm{AP}$}

Here we outline the method we use to quantify the fidelity of the adiabatic transport protocol. Our strategy will be to truncate the Hamiltonian of the system to the relevant $3\times 3$ block. We then map this onto a $2\times 2$ Hamiltonian and translate it into the adiabatic frame to estimate the fidelity.

We start with the basic Hamiltonian $H_{s-c-t}$ for the source-chain-target system defined in Equation \ref{eq:Hamiltonian_source_chain_target}. As described by Oh \textit{et al.}, $H_{s-c-t}$ is block-diagonal, consisting of four blocks. Only two blocks play a role in adiabatic transport, as described in Section \ref{sec:adiabatic_transport}, where one transports the $\lvert{\downarrow_s}\rangle$ component and the other transports the $\lvert{\uparrow_s}\rangle$ component of the source qubit. When $\Delta B_z=0$, these two blocks are identical apart from the Zeeman energy. The Zeeman energy can be ignored, as it is simply an identity offset to the diagonal elements of either block. Either adiabatic transport block is defined by the following Hamiltonian.

\begin{equation}
H_\mathrm{AP-3}(t)=  \frac{1}{4}\begin{pmatrix}
J_t - J_s & 2J_s                      & 0 \\
2J_s                     & -J_s - J_t  & 2J_t \\
0                        & 2J_t                      & J_s - J_t
\end{pmatrix}
\end{equation}

The basis states of $H_\mathrm{AP-3}$ are given by $\{\lvert \downarrow_s \uparrow_c \uparrow_t  \rangle$, $\lvert \uparrow_s \downarrow_c \uparrow_t  \rangle , \lvert \uparrow_s \uparrow_c \downarrow_t  \rangle\}$ for the spin-up block and $\{\lvert \uparrow_s \downarrow_c \downarrow_t  \rangle$, $\lvert \downarrow_s \uparrow_c \downarrow_t  \rangle , \lvert \downarrow_s \downarrow_c \uparrow_t  \rangle\}$ for the spin-down block. Arbitrarily, we analyze the fidelity for transporting the $\lvert{\uparrow_s}\rangle$ source state. The eigenvectors of $H_\mathrm{AP-3}$  are given by \cite{Oh2013pra}

\begin{subequations}
	\begin{equation}
	\lvert\Phi_0(t)\rangle = \frac{1}{\sqrt{3}}
	\begin{pmatrix}
	1\\
	1\\
	1
	\end{pmatrix},
	\end{equation}
	\begin{equation}
	\lvert\Phi_{\pm}(t)\rangle = \frac{1}{\sqrt{N_{\pm}}}
	\begin{pmatrix}
	\mathrm{sin}(\zeta)\\
	-\mathrm{cos}(\zeta) \pm \sqrt{q}\\
	\mathrm{cos}(\zeta) - \mathrm{sin}(\zeta) \mp\sqrt{q}
	\end{pmatrix},
	\end{equation}
	\label{eq:Eigen_vectors_3by3_Hamiltonian}
\end{subequations}

where $\mathrm{tan}(\zeta) = \left(J_s(t)/J_t(t)\right)$, $q =  1 - \mathrm{sin}(\zeta)\mathrm{cos}(\zeta)$, and $N_{\pm} = \mp 2\left(2\mathrm{cos}(\zeta) - \mathrm{sin}(\zeta)\right)\sqrt{q} + 4q$. For adiabatic transport, the system is initialized in the $\lvert\uparrow_s S_{ct}\rangle$ state, which is $\lvert\Phi_{-}(0)\rangle$.

Recall that the eigenenergies of the $\lvert \uparrow\rangle$-transport and $\lvert \downarrow \rangle$-transport blocks are plotted in Figure \ref{fig:APEigStates}. We see that in each block, two states anti-cross whereas the energy of one state is constant, suggesting that $H_\mathrm{AP-3}(t)$ can be truncated to a $2\times 2$ Hamiltonian. For this, we write $H_\mathrm{AP-3}$ in the basis of the eigenstates at $t = T_\mathrm{AP}/2$, which can be obtained by substituting $J_s = J_t$ into Equation \ref{eq:Eigen_vectors_3by3_Hamiltonian}. The new basis states are given by

\begin{equation}
\lvert\Phi_0^{'}\rangle = \frac{1}{\sqrt{3}}
\begin{pmatrix}
1\\
1\\
1
\end{pmatrix},
\lvert\Phi_+^{'}\rangle = \frac{1}{\sqrt{2}}
\begin{pmatrix}
1\\
0\\
-1
\end{pmatrix},
\lvert\Phi_-^{'}\rangle = \frac{1}{\sqrt{6}}
\begin{pmatrix}
1\\
-2\\
1
\end{pmatrix}.
\end{equation}

The Hamiltonian $H_\mathrm{AP-3}$ in this basis, $\{\lvert\Phi_0^{'}\rangle, \lvert\Phi_+^{'}\rangle, \lvert\Phi_-^{'}\rangle\}$, is

\begin{equation}
H_\mathrm{AP-3}^{'}(t)=  \frac{1}{4}\begin{pmatrix}
J_t + J_s & 0                      & 0 \\
0                     & 0  & \sqrt{3}(J_t - J_s) \\
0                        & \sqrt{3}(J_t - J_s)                      & -2J_s - 2J_t
\end{pmatrix}.
\end{equation}

As expected, $H_\mathrm{AP-3}^{'}(t)$ is block-diagonal. The initialized state of the system is $\lvert\uparrow_s S_{ct}\rangle = (1/2)\lvert\Phi_+^{'}\rangle - (\sqrt{3}/2)\lvert\Phi_-^{'}\rangle$. As the initial population of the $\lvert\Phi_0^{'}\rangle$ is zero, we truncate $H_\mathrm{AP-3}^{'}(t)$ to the lower $2\times 2$ block, spanned only by $\lvert\Phi_+^{'}\rangle$ and $\lvert\Phi_-^{'}\rangle$. Adding $(J_s+J_t)I_2$, we obtain

\begin{equation}
H_\mathrm{AP-2}^{'}(t)=  \frac{1}{4}\begin{pmatrix}
J_t+J_s &  \sqrt{3}(J_t - J_s)\\
\sqrt{3}(J_t - J_s) & -(J_t + J_s)
\end{pmatrix}.
\end{equation}

Recall that $J_s+J_t = J_\mathrm{max}$ and $J_t - J_s = J_\mathrm{max}(1-2t/T_\mathrm{AP})$. We thus complete our mapping onto a spin in a magnetic field, obtaining

\begin{equation}
H_\mathrm{AP-2}^{'}(t) = \varepsilon\frac{\sigma_z}{2} + \Delta_\mathrm{AP}(t)\frac{\sigma_x}{2},
\label{eq:Mapping AP two level system}
\end{equation}

where $\varepsilon = {J_\mathrm{max}}/{2}$ and $ \Delta_\mathrm{AP}(t)  = (\sqrt{3}J_\mathrm{max}/2)\left(1- 2{t}/{T_\mathrm{AP}}\right)$. Note that the initialized state, $\lvert\uparrow_s S_{ct}\rangle$ is the lower-energy eigenstate of $H_\mathrm{AP-2}^{'}(0)$.

With the mapping complete, we then move to calculating the fidelity of the adiabatic protocol. For this, we translate $H_\mathrm{AP-2}^{'}(t)$ into the adiabatic frame, by invoking the same operation used in Equation \ref{eq:Singlet_Initialization_adiabatic_frame}. The Hamiltonian in the adiabatic frame $H_\mathrm{AP-2}^{A}(t)$ can be simplified as

\begin{equation}
H_\mathrm{AP-2}^{A}(t) = \sqrt{\varepsilon^2 + \Delta_\mathrm{AP}(t)^2}\frac{\sigma_z}{2} - \hbar \frac{d\chi_\mathrm{AP}}{dt}\frac{\sigma_y}{2},
\label{eq:Mapping AP adiabatic_frame}
\end{equation}

where $\mathrm{tan}(\chi_\mathrm{AP}) = \Delta_\mathrm{AP}(t)/\varepsilon$. The eigenvectors of $H_\mathrm{AP-2}^{A}(t)$ are given by

\begin{subequations}
\begin{equation}
\lvert\Phi_{1_\mathrm{AP}}(t)\rangle =
  \begin{pmatrix}
    \mathrm{cos}(\alpha_\mathrm{AP})\\
    -i\mathrm{sin}(\alpha_\mathrm{AP})
  \end{pmatrix},
\end{equation}
\begin{equation}
\lvert\Phi_{2_\mathrm{AP}}(t)\rangle =
  \begin{pmatrix}
    \mathrm{sin}(\alpha_\mathrm{AP})\\
    i\mathrm{cos}(\alpha_\mathrm{AP})
  \end{pmatrix},
\end{equation}
\label{eq:Mapping_AP_adiabatic_frame_eigen_state}
\end{subequations}

where $\mathrm{tan}(2\alpha_\mathrm{AP}) = \hbar\frac{d\chi_\mathrm{AP}}{dt}/\sqrt{\varepsilon^2 + \Delta_\mathrm{AP}(t)^2}$. We employ the same technique used in Appendix \ref{appendix:SPfid} to estimate the fidelity of the protocol. We consider the dynamics in the adiabatic frame and assume the adiabatic limit, where the projection of the instantaneous state onto the eigenstate remains constant. The diameter of the `circle of precession,' again, is determined by the initial eigenstates in the adiabatic frame, which is at an angle $\alpha_\mathrm{AP}(0)$ from the $z$-axis. The eigenstate at the end of the protocol is oriented at an angle $\alpha_\mathrm{AP}(T_\mathrm{AP})$ from the $z$-axis. From Equation \ref{eq:Mapping_AP_adiabatic_frame_eigen_state}, we see that $\alpha_\mathrm{AP}(T_\mathrm{AP}) = \alpha_\mathrm{AP}(0)$. Therefore, at the end of the protocol, the circle of precession is at exactly the same position as at $t=0$. Recall that the fidelity of the adiabatic protocol is the projection-squared of the final state onto the $z$-axis. The `circle of precession' touches the $z$ axis, and hence we obtain resonances in the transport fidelities as a function of $T_\mathrm{AP}$ in Figure \ref{fig:APFidelity_JmaxTAP}. To obtain the worst case fidelity, however, we use the opposite point on the circle, which has the maximum angular deviation from the $z$-axis. This yields an expression for the envelope of transport errors given by

\begin{equation}
1 - F_\mathrm{AP} \approx \mathrm{sin}^2(2\alpha_\mathrm{AP}(0)) = \frac{3}{3 + \left(2  J_\mathrm{max}T_\mathrm{AP}/\hbar\right)^2}
\end{equation}

Note that this expression perfectly aligns with the numerical simulations in the adiabatic regime ($J_\mathrm{max}T_\mathrm{AP}/h \gg 1$) in Figure \ref{fig:APFidelity_JmaxTAP}.

\subsubsection{Exchange tunability errors: ${J_\mathrm{min}}/{J_\mathrm{max}}$}

Here we derive an expression for the transport fidelity limited by `spin-leakage,' as defined in Section \ref{sec:AP_Jmin}. We set $\Delta B_z = 0$ for simplicity, such that the $\lvert\uparrow\rangle$ and $\lvert\downarrow\rangle$ adiabatic passages are equivalent. In Section \ref{sec:AP_Jmin}, in the example of transporting the $\lvert\uparrow_s\rangle$ state, we have considered the system to be initialized in the $\lvert \uparrow_s S_{ct}\rangle$ state. At $t=T_\mathrm{AP}$, the system is transported to the $\lvert S_{sc}\uparrow_t\rangle$ state, which is not an eigenstate if $J_\mathrm{min} \neq 0$. To quantify the error due to the resulting precession, we express $\lvert S_{sc}\uparrow_t\rangle$ as a superposition of the eigenstates of $H_\mathrm{AP-3}$, given by

\begin{equation}
\lvert S_{sc}\uparrow_t\rangle = c_+\lvert \Phi_+(T_\mathrm{AP}) \rangle + c_-\lvert \Phi_-(T_\mathrm{AP})\rangle,
\label{eq:Singlet_components_Jmin}
\end{equation}

where $c_\pm^2 = \left[ \mp\left(2\cos\left(\zeta_0\right) - \sin\left(\zeta_0\right)\right) + 2\sqrt{q}\right]/{4\sqrt{q}}$ with $\tan(\zeta_0) = J_\mathrm{min}/J_\mathrm{max}$. For $J_\mathrm{min}/J_\mathrm{max} < 1$, $c_- > c_+$. The precession frequency is $\sim J_\mathrm{max}/h$. The maximum error is when the state rotates by an angle $\pi$ around the eigenstate, such that the state of the system becomes $c_+\lvert \Phi_+(T_\mathrm{AP}) \rangle - c_-\lvert \Phi_-(T_\mathrm{AP})\rangle$. This worst-case leads to a fidelity given by $F_\mathrm{AP} =  \lvert c_+^*c_+ - c_-^*c_- \rvert ^2 $, which is the projection onto the  $\lvert S_{sc}\uparrow_t\rangle$ state. This can be simplified to

\begin{equation}
1 - F_\mathrm{AP} = \frac{3}{4}\times\frac{\left(J_\mathrm{min}/J_\mathrm{max}\right)^2}{ \left(J_\mathrm{min}/J_\mathrm{max}\right)^2 - \left(J_\mathrm{min}/J_\mathrm{max}\right) + 1}.
\label{eq:Fidelity_Jmin_simplified}
\end{equation}

In the limit $J_\mathrm{min}/J_\mathrm{max} \ll 1$, the transport error takes the form

\begin{equation}
1 - F_\mathrm{AP} \approx \frac{3}{4}\times(J_\mathrm{min}/J_\mathrm{max})^2.
\end{equation}

\bibliographystyle{ieeetr}

\end{document}